\documentclass[epj,nopacs]{svjour}

\def\bea{\begin{eqnarray}}
\def\eea{\end{eqnarray}}
\def\beq{\begin{equation}}
\def\eeq{\end{equation}}

\def\leff{\mathcal{L}_{\rm eff}}

\usepackage{graphics}
\usepackage{epsf}
\begin{document}
\title{Chiral order and fluctuations in multi-flavour QCD}
\author{S.~Descotes-Genon\inst{1} \and L.~Girlanda
\thanks{
Present address: ECT$^*$, Villa Tambosi, Strada delle Tabarelle 286, I-38050
 Trento, Italy.}
\inst{2}
\and J.~Stern\inst{3}}
%
\institute{Department of Physics and Astronomy,
University of Southampton, Southampton SO17 1BJ, U.K.\\
\email{sdg@hep.phys.soton.ac.uk} \and
Dipartimento di Fisica, Universit\`a di Padova and
INFN, Via Marzolo 8, I-35131 Padova, Italy\\ \email{luca.girlanda@pd.infn.it}
\and
Groupe de Physique Th\'eorique, Institut de Physique Nucl\'eaire, F-91406
Orsay-Cedex, France\\ \email{stern@ipno.in2p3.fr} }
\date{}
%
\abstract{
Multi-flavour $(N_f \ge 3)$ Chiral Perturbation Theory ($\chi$PT)
may exhibit instabilities due to vacuum fluctuations of sea $\bar q q$-pairs.
Keeping the fluctuations small would require a very precise fine-tuning 
of the low-energy constants $L_4(\mu)$ and $L_6(\mu)$ to 
$L_4^{\rm crit}(M_\rho) = - 0.51 \cdot 10^{-3}$, 
$L_6^{\rm crit}(M_\rho) = - 0.26 \cdot 10^{-3}$. A small deviation
from these critical values -- like the one suggested by the phenomenology 
of OZI-rule violation in the scalar channel -- is amplified by
huge numerical factors inducing large effects of vacuum fluctuations.
This would lead in particular to a strong $N_f$-dependence of Chiral
Symmetry Breaking ($\chi$SB) 
and a suppression of multi-flavour chiral order parameters. 
A simple resummation is shown to cure 
the instability of $N_f\ge 3$ $\chi$PT, but it modifies the standard 
expressions of some $O(p^2)$ and $O(p^4)$ low-energy parameters in terms of 
observables. On the other hand, for $r= m_s/m > 15$, the two-flavour 
condensate is not suppressed, due to the contribution induced by 
massive vacuum $\bar ss$-pairs. Thanks to the latter, the standard 
two-flavour $\chi$PT is protected from multi-flavour instabilities
and could provide a well-defined expansion scheme in powers of non-strange
quark masses.
} 
\titlerunning{Chiral order and fluctuations in multi-flavour QCD}
\authorrunning{S.~Descotes, L.~Girlanda and J.~Stern}

\maketitle

\section{Introduction} \label{sec:intro}

Understanding  Chiral Symmetry Breaking ($\chi$SB) in low-energy QCD still
deserves both phenomenological and theoretical effort. First, there is a
growing need to identify and to separate non-perturbative
QCD effects from possible
manifestations of ``New Physics'' in experimental tests of the Standard
Model (e.g.  weak matrix elements, $\epsilon'/\epsilon$,
$(g - 2)_\mu$\ldots). Furthermore, the subject has its own theoretical
interest. Vector-like gauge theories such as QCD formulated in a large
Euclidean box allow a particularly attractive interpretation of spontaneous
$\chi$SB in terms of the lowest modes of the Dirac operator
averaged over all gluon configurations~\cite{dirac}.
In QCD-like theories, some characteristic properties of
the Dirac spectrum have been proven~\cite{VW1,VW2} and possible consequences
for chiral order parameters have been conjectured~\cite{DGS}.
More generally, this approach to $\chi$SB suggests an analogy with
disordered systems of higher dimensionality  ($d = 4$) emphasising
notions such as the average
and the fluctuation of the density of small Dirac eigenvalues, as well as
the transport properties (e.g. conductivity)~\cite{S}.
Finally, the cornerstone
of all this investigation is Chiral Perturbation Theory
($\chi$PT)~\cite{GL1,GL2} which provides a systematic link
between theoretical characteristics of $\chi$SB
(described by order and fluctuation parameters) on one hand and
observable properties of Goldstone bosons (masses, decay constants,
scattering amplitudes, decay form-factors\ldots) on the other hand.

              During the last few years both theoretical and experimental
progress was achieved along these lines. It has been suggested that order
parameters of $\chi$SB, in particular the quark-antiquark condensate
$\langle\bar qq\rangle$, could strongly depend on the number $N_f$ of
light flavours~\cite{nfdep,nfdeplat}. 
As $N_f$ increases, $\langle\bar qq\rangle$
as well as the Goldstone boson coupling
$F_\pi$ are gradually suppressed, due to the paramagnetic behaviour of
Dirac eigenvalues and to increasing fluctuations of the density of
states~\cite{VW2,DGS}. This effect is induced by light-quark loops and
it cannot be detected in quenched lattice simulations. Actually, there are
two kinds of paramagnetic effects generated by loops of sea-quarks
which are both of the same origin~\footnote{A sea-quark
loop is one with no external source attached to it.}:
the massless loops suppress chiral order parameters whereas
the massive sea quark pairs enhance them, as long as their mass is of order
$\Lambda_{QCD}$ or smaller. In Nature, this last remark merely concerns
the strange quark, whose mass is slightly below $\Lambda_{QCD}$.
The abundance of strange quark-antiquark pairs in the vacuum
can thus lead to a different behaviour of two-flavour
($m_s \sim \Lambda_{QCD}$) and three-flavour ($m_s = 0$) chiral dynamics.
Such difference would be characterised by a nonnegligible vacuum correlation
between strange and non-strange quark pairs, implying in turn large
$1/N_c$ corrections and violation of the OZI-rule in the scalar channel.
The latter is actually observed~\cite{scalar} and a strong variation of
$\chi$SB between $N_f = 2$ and $N_f = 3$ has been indeed reported 
on the basis of sum-rule studies~\cite{Bachir1,Bachir2,D} using
as input available information about the scalar sector $0^{++}$ .

Such a possibility should now be considered in the light of the new
experimental information on low-energy $\pi\pi$ scattering
which has been recently published~\cite{E865} and
analysed~\cite{ACGL,pipiCGL,CGLNPB,CGLPRL,DFGS}.
The outcome of these analyses shows that, in the
presence of massive $\bar ss $ pairs in the vacuum, the two-flavour
condensate $\langle\bar{u}u\rangle$ is large and
dominates the $ SU(2) \times SU(2)$ symmetry-breaking
effects~\cite{CGLPRL,DFGS}. Accordingly, the standard two-flavour $\chi$PT
expansion in powers of $m_u$ and $m_d$~\cite{GL1} should be expected
to converge rather well. On the other hand, large vacuum fluctuations of
$\bar qq$-pairs would result into a large difference between
$N_f=2$ and $N_f=3$ condensates, destabilising the three-flavour expansion.
Indeed, a detailed $SU(3) \times SU(3)$ analysis of Goldstone boson
masses and decay constants within the standard two-loop $\chi$PT~\cite{ABT}  
has revealed an anomalously large $O(p^6)$ contribution to $M_\pi^2$, 
depending on a fine tuning of the LEC's $L_4(\mu)$ and $L_6(\mu)$ -- which 
precisely reflect vacuum fluctuations. 
The purpose of this paper is to show that the 
difference in the chiral behaviour of two-flavour and multi-flavour 
QCD described above could be naturally explained in terms of the 
interplay between vacuum fluctuations (of small Dirac eigenvalues) 
and chiral order (described by order parameters such as $\langle\bar qq\rangle$) .

                  We start by considering Ward identities and low-energy
theorems for the two-point functions 
$\langle D^a D^b\rangle$ and $\langle V^a V^b - A^a A^b\rangle$,
where $V^a$, $A^a$ and $D^a$ are (charged) vector currents, axial currents
and the divergences of the latter, respectively.
We write these identities in a form reminiscent of the
$\chi$PT expansion of $M^2_P F^2_P$ and $F^2_P$ ($P=\pi,K,\eta$), including
explicitly the leading and next-to-leading orders in powers of quark
masses~\cite{GL2} and collecting (\emph{not} neglecting) all remaining orders
into well-defined ``remainders''. We refer to Ward Identities
written in this way as ``mass and decay constant identities''. 

We then show that there exist two exact nonlinear
relations between the order parameters $\langle\bar qq\rangle$ and
$F^2_P$ in the chiral limit $SU(N_f) \times SU(N_f)$
with $N_f \ge 3$ and two ``fluctuation
parameters'' which are defined in terms of the standard LEC's $L_6(\mu)$
and $L_4(\mu)$: in these relations, all the effects of higher $\chi$PT orders
are absorbed into a finite multiplicative renormalisation of order and
fluctuation parameters. The deviation from 1 of the corresponding
renormalisation constants (rescaling factors) remains under
control to the extent that the NNLO
remainders in mass and decay constant identities are small. The $\chi$PT series
is reproduced in the limit of small fluctuation parameters, implying
a very precise fine tuning of $L_6(\mu)$ and $L_4(\mu)$. Otherwise (and
in particular for large fluctuation parameters), the multi-flavour chiral
condensate is suppressed, and the standard $\chi$PT interpretation of mass and
decay constant identities breaks down. This need not affect the overall
convergence of the $\chi$PT series. We actually expect that for the physical
value of $m_s$ the multi-flavour $\chi$PT still makes sense globally. 
In this case the instability caused by large fluctuations and by the 
suppression of the quark condensate can be cured by a simple ``resummation'' 
which amounts to replacing
the perturbative solution of the non-linear relations between order and
fluctuation parameters by their exact algebraic solution. This merely modifies
the standard way of expressing the parameters of the effective Lagrangian
in terms of observable quantities~\cite{GL2}. 

The multi-flavour mass and decay constant identities can be
further used to define the two-flavour order parameters,
by taking the limit $m_u,m_d \to 0$ but keeping $m_s$ fixed
at its physical value.
In this way we show that the two-flavour condensate and decay constant are
not affected by the large fluctuations which suppress the multi-flavour
condensate. In addition, this allows one to discuss the
connection with two-flavour $\chi$PT and the new $\pi\pi$
scattering data. 

The plan of this article is the following. In Sec.~\ref{sec:2}, we
discuss the impact 
of loops of massive sea-quarks (typically, $s\bar{s}$-pairs) on
the pattern of $\chi$SB. 
We introduce in Sec.~\ref{sec:model} the mass and decay constant
identities for a  generic $N_f \ge 3$, and show that they lead to an
exact system of relations between order and fluctuation parameters,
presented in Sec.~\ref{sec:4}.
The general properties of
this system are then discussed in the light of the positivity and (conjectured)
paramagnetic inequalities that the order and fluctuation parameters have to
obey (Sec.~\ref{sec:5}). Sec.~\ref{sec:6} is devoted to the study (in the plane of
fluctuation parameters) of the critical line where the symmetry is restored,
i.e. both the condensate and the decay constant vanish.
Sec.~\ref{sec:7} briefly summarises
properties of the large-$N_c$ limit in which  fluctuations are suppressed.
Then in Sec.~\ref{sec:8} we discuss the opposite limit of large
fluctuations. A possible realisation of the latter can be interpreted
as a limit of large $N_f$.
It is shown that the limit of large fluctuations
is in principle different from the symmetry
restoration limit: despite the continuous vanishing of the multi-flavour
quark condensate, the decay constant stays non-zero.
Sec.~\ref{sec:9} deals with the $SU(2)\times SU(2)$ chiral limit: we check once
more that in the large-fluctuation limit the two-flavour condensate remains
non-zero and the two-flavour Gell-Mann--Oakes--Renner relation is
approximately obeyed, and we discuss briefly the recent results
on $\pi\pi$ scattering. The conclusion and a few appendices close the paper.

\section{Role of the mass of sea quarks}
\label{sec:2}

                      When defining a chiral limit or a chiral order parameter,
it should be stated which fermions are taken massless and which quarks in
the sea are left massive. The simplest situation is the one with just $N$
massless fermions $\psi_1\ldots \psi_N$,   
(i.e. $m_1 = m_2 =\ldots m_N = m \to 0$) and no other massive fermions left.
The condensate of this purely massless theory is defined as:
\begin{equation}\label{condmassless}
\delta_{ij} \sigma(N) = - \lim_{m \to 0} \langle \bar \psi_i \psi_j \rangle
         = - \langle \bar \psi_i \psi_j \rangle_N\,.
\end{equation}
The expression of the condensate in terms of eigenvalues $\lambda_n$ of the
Euclidean Dirac operator $ \gamma_{\mu} D_\mu[G]$ defined in a box
$L \times L \times L \times L$ with periodic boundary conditions (up to a
gauge transformation) is well known as the Banks-Casher formula~\cite{dirac}.
Formally, it can be written as:
\begin{equation}\label{bankcash}
\sigma(N) = \lim \frac{1}{V}\ll \sum_n \frac{m}{m^2 + \lambda^2_n [G]}\gg \,.
\end{equation}
where $\lim$ means taking $V \to \infty$ first and $m \to 0$ afterwards.
The average $\ll$  $\gg$ over Euclidean gluon configurations involves the
$N$-th power of the fermion determinant $\Delta^N(m,G)$. Since for
$m \to 0$, $V \to \infty$ only the smallest eigenvalues contribute in
Eq.~(\ref{bankcash}), it is conceivable that the main $N$-dependence
merely arises from the infrared part of the determinant          
$\Delta(m,G) = \Delta_{IR}\Delta_{UV}$, defined as:
\begin{equation}\label{infradet}
\Delta_{IR} (m,G) = m^{|\nu|} \prod_{n<K} 
\frac{m^2 + \lambda^2_n}{\mu^2 + \omega^2_n} \,,
\end{equation}
where $\nu$ is the winding number of the gluon
configuration $G$  and   $K(\Lambda,G)$ is an integer corresponding to a    
cutoff $\Lambda$ such that
$\lambda_K = \Lambda$. Positive eigenvalues are ranked in ascending order
$\lambda_1 < \lambda_2 <\ldots <\lambda_K = \Lambda$. The numbers $\omega_n$
are defined by the Vafa-Witten bound~\cite{VW2} for the Dirac eigenvalues:
\begin{equation}\label{VW}
\lambda_n < C \left(\frac{n}{V}\right)^{1/d}\equiv \omega_n\,.
\end{equation}
The existence of such a uniform bound independent of gauge field
configurations is a specific property of QCD-like gauge theories. It
implies $(\Delta_{IR})^{N+1} < (\Delta_{IR})^N $ for any finite cutoff $K$
and for $m<\mu$. In the chiral limit $V \to \infty , m \to 0$ one can then 
expect the paramagnetic inequality:
\begin{equation}\label{masslesspara}
\sigma(N + 1)\le \sigma( N )\, .
\end{equation}
                        In the real world, the situation is slightly more
complicated due to the role of massive virtual quark pairs which may be
present in the vacuum. Notice that massive and massless pairs have different
chiral transformation properties and do not affect the chiral structure of
the vacuum in the same way.  In QCD one deals with 
a hierarchy of quark masses:
\begin{equation}
m_u < m_d \ll m_s \ll m_c < m_b \ll m_t\,.
\end{equation}
Some of them ($u , d , s$) can be considered as light compared to the scale
$\Lambda_H \sim 1$ GeV at which the masses of the first
bound states non-protected by chiral symmetry occur.
$\Lambda_H \sim 4 \pi F_\pi $ is the reference scale in $\chi$PT expansions
in powers of $p/\Lambda_H$. A different question can be asked in connection
with the structure of the vacuum. Some quarks ($c,b,t$)
can be considered as too heavy to form abundant vacuum pairs. In this case, the
characteristic scale is not $\Lambda_H$ but the lower scale $\Lambda_{QCD}$.
The reason is that we are not interested in the production of massive
hadrons but in the creation of massive virtual $\bar qq$ pairs. The latter
will be most probable if the quark mass is of order $\Lambda_{QCD}$ or
slightly lower (if $m \to 0$, the chiral properties of the corresponding pairs
tend smoothly to the massless case).

This reasoning already singles out the strange
quark among all six quarks we know. Its mass $m_s \sim 160$ MeV
(at $\mu = 1$ GeV)~\cite{ms} is sufficiently low compared to $\Lambda_H$         
to legitimate a $SU(3) \times SU(3)$ chiral expansion. 
On the other hand, $m_s$ is sufficiently close to $\Lambda_{QCD}$
to expect a significant presence of massive $\bar ss$-pairs in the vacuum.

We define the $SU(N_f) \times SU(N_f)$ chiral limit
in QCD by taking the first $N_f$ quarks as massless and keeping the remaining
masses at their physical value. In practice, one can consider such a limit
for $N_f=2$ or for $N_f=3$. The corresponding order parameters will be
functions of the remaining non-zero masses. For instance, the two-flavour
condensate is defined as:
\begin{equation}\label{Sigma2}
\Sigma(2 , m_s, \ldots ) = - \lim_{m_u , m_d\to 0} \langle \bar u u \rangle \,,
\end{equation}
and it is a function of $m_s$ as well as of the heavy-quark masses denoted in
Eq.~(\ref{Sigma2}) by the ellipsis. $u(x)$ stands for the lightest ($u$) quark
field, and it can equivalently be replaced in Eq.~(\ref{Sigma2}) by the
$d$-quark, but not by the $s$-quark field.  The three-flavour condensate 
is then defined as:
\begin{equation}\label{Sigma3}
\Sigma( 3 ,\ldots ) = \lim_{m_s \to 0} \Sigma(2, m_s, \ldots )\,,
\end{equation}
One  expects that the effect of heavy-quark masses
$m_c$, $m_b$, $m_t$ on $\Sigma(2)$ and $\Sigma(3)$ remains small and could
be eventually estimated. For simplicity, we shall neglect all effects of 
heavy-quark masses in the sequel. In this approximation, $\Sigma(3)$ coincides
with $\sigma(3)$ defined in Eq.~(\ref{condmassless}). It is a clean probe of the chiral
structure of the vacuum of QCD with nothing but three massless quarks:
once one sets $m_s = 0$, there is no more massive
quark left which would be sufficiently light to pollute the vacuum
$|\Omega\rangle_3$.

The situation is rather different in the
two-flavour chiral limit $m_u = m_d = m \to 0 $, keeping the strange-quark
mass $m_s$ at its physical value. Since $m_s$ is not very large compared
to $\Lambda_{QCD}$ and vacuum is polluted by massive $\bar ss$ pairs, it is
difficult to relate  $\Sigma(2)$ to the genuine
condensate $\sigma(2)$ characteristic of the theory
with nothing but two massless quarks. This situation occurs as long as
$m_s$ remains of order $\Lambda_{QCD}$. One can gain more
insight into the $m_s$-dependence of $\Sigma(2)$ from the formula in
Euclidean space:
\begin{equation}\label{strangesee}
\Sigma(2)=-\frac{\langle \bar uu \exp[-m_s\int\bar ss]\rangle_3}
{\langle\exp[-m_s\int\bar ss]\rangle_3}\,,
\end{equation}
where on the right hand side the expectation value is taken with respect to
the vacuum of the $SU(3) \times SU(3)$ invariant theory. It is seen that in 
the absence of correlations between the strange and non-strange $\bar qq$ pairs,
Eq.~(\ref{strangesee}) implies $\Sigma(2) = \Sigma(3)$, in agreement with
large-$N_c$ expectations. For $m_s \ll \Lambda_H$, Eq.~(\ref{strangesee})
can be rewritten as:
\begin{equation}\label{23}
\Sigma(2) = \Sigma(3)  +  m_s \bar Z^s_1  + \ldots 
\end{equation}
where $\bar Z^s_1$ denotes the connected correlator of $\bar ss$- and 
$\bar uu$-pairs:
\begin{equation}\label{sucorr}
\bar Z^s_1 = \lim_{m \to 0} \int dx \ \langle \bar ss(x) \bar
uu(0)\rangle_{\rm con}\,,
\end{equation}
Notice that in the limit $m_{u,d} \to 0$, $\bar uu$ can be equally replaced
by $(\bar uu + \bar dd)/2$. The correlator~(\ref{sucorr}) measures the
violation of the OZI rule in the isoscalar scalar (i.e. vacuum) channel
and it can be estimated using the experimental information now available in
this channel~\cite{Bachir1,Bachir2,D}. It is related to the standard $O(p^4)$ LEC
$L_6(\mu)$ and it turns out to be larger than what is expected 
on the basis of large-$N_c$ considerations~\cite{GL2}.

                      It is useful to express $\Sigma(2)$, $\Sigma(3)$ and
the correlator $\langle(\bar ss) (\bar uu)\rangle$ 
in terms of the eigenvalues $\lambda_n$
of the Euclidean Dirac operator. Neglecting heavy quarks, both $\Sigma(3)$
and $\Sigma(2)$ concern the theory with the same total number of fermions:
in $\Sigma(3)$, one sets $ m_u = m_d = m_s = m \to 0$, whereas in
$\Sigma(2)$ $m_u = m_d = m  \to 0$ but $m_s \sim \Lambda_{QCD}$ is held
fixed. Hence, the corresponding Banks-Casher formula becomes~\cite{dirac}:
\begin{equation}\label{3-2}
\Sigma(N_f) = \lim \frac{1}{V} \ll \sum_n \frac{m}{m^2 + \lambda^2_n}
\gg_{N_f}\,,
\end{equation}
where the only difference lies in the determinant inserted in the average
over gluon configurations: $\Delta^3(m,G)$ for $\Sigma(3)$ and
$\Delta^2(m,G) \Delta(m_s,G)$ in the case of $\Sigma(2)$. Comparing
the corresponding infrared parts Eq.~(\ref{infradet}) which are expected to
dominate in Eq.~(\ref{3-2}), one has:
\begin{eqnarray}\label{paradet}
&& m^{3|\nu|} \prod_{n<K}
\left(\frac{m^2+\lambda^2_n}{\mu^2+\omega^2_n}\right)^3 \nonumber \\
&\le&
m^{2|\nu|}m_s^{|\nu|} \prod_{n<K}\frac{(m^2+\lambda^2_n)^2(m_s^2+\lambda^2_n)}           
{(\mu^2+\omega^2_n)^3} \,,
\end{eqnarray}
as long as $m \to 0$ and $m_s \sim \Lambda_{QCD}$. This suggests that the
paramagnetic inequality~(\ref{masslesspara}) holds even in the presence of
massive strange quarks in the sea:
\begin{equation}\label{paramass}
\Sigma ( 2 ) \ge \Sigma ( 3 )\,.
\end{equation}
The conjectured inequality Eq.~(\ref{paramass}) will play an important role in
the sequel.

              It remains to express the correlator Eq.~(\ref{sucorr}) in terms
of Dirac eigenvalues; see e.g. Ref.~\cite{DGS}. One has
\begin{equation}\label{flu}
\bar Z^s_1 = \lim \frac{1}{V} \ll \sum_{nk} \frac{m}{m^2+\lambda^2_n}
               \frac{m_s}{m_s^2+ \lambda^2_k} \gg_{\rm con}    \,.
\end{equation}
For small $m$ and $m_s$, only small Dirac eigenvalues contribute and the
expression (\ref{flu}) measures their correlations. For $m_s \sim m$,
$\bar Z^s_1$ describes the fluctuations of the density of states
$ \rho(\lambda) = \sum_n \delta (\lambda-\lambda_n)$ near
$\lambda \sim 0$. The positivity of Eq.~(\ref{flu}) is in agreement with the
paramagnetic inequality Eq.~(\ref{paramass}).      
In this paper, we investigate the possibility
that multi-flavour QCD vacuum behaves as a strongly correlated 
fermion--anti-fermion system characterised by large fluctuations
$\bar Z^s_1$ of the density of small Dirac eigenvalues.

                 Let us mention without proof that the present discussion
can be easily extended to the order parameter:
\begin{equation}\label{F}
F^2 (N_f) = \lim_{m_1\ldots m_{N_f} \to 0} F^2_\pi\,,
\end{equation}
defined as before by taking the first $N_f$ quarks massless and leaving the
remaining quarks at their physical masses. The dependence of these order
parameters on  $N_f$ and on the sea-quark masses is qualitatively similar to
that of the quark condensates.

\section{$SU(N_f)\times SU(N_f)$ chiral symmetry for  $N_f\geq 3$} 
\label{sec:model}

We consider QCD with $N_f=n+2$ light flavours which is very like our
actual three flavour QCD equipped with $n$  copies of the strange
quark degenerate in mass: $u$ and $d$ quarks with ultralight
degenerate masses $m_u=m_d=m$  and $n$ ``strange'' copies $s_1...s_n$
with a common mass $m_s \gg m$ but still light compared to the QCD
scales.

In the chiral limit $m = m_s =0$, the
$SU(2+n)\times SU(2+n)$ chiral symmetry is assumed  to be spontaneously
broken down to the diagonal subgroup $SU_V(n+2)$. This symmetry is
explicitly broken by the mass difference $m_s-m$ to $SU(2)\times SU(n)$,
which is the exact symmetry of our problem. We have
$(n+2)^2 - 1 = n^2 + 4n +3 $ pseudo-Goldstone bosons: 3 pions of mass $M_\pi$,
$4n$ kaons ($\bar us_i , \bar ds_i$ for $i=1\ldots n$, plus conjugates) of common
mass $M_K$, the $\eta$-meson  with structure
$\lambda_\eta = (1+2/n)^{-1/2}{\rm diag}[ 1,1 ; - 2/n \ldots -2/n]$ and mass $M_\eta$,
and finally the extra $n^2-1$ $\bar s_is_j$ states whose mass will be denoted
as $M_X$. 

This framework will be used to discuss the response
of order parameters of the chiral symmetry $SU_L(N_f)\times SU_R(N_f)$
to large vacuum fluctuations both for $N_f \ge 3$ and for $N_f=2$.
We will see that these two cases behave rather differently.
It will prove useful to keep a generic value for $N_f$, although our
main interest will be on the case $N_f=3$ and all our conclusions
apply to this case.

\subsection{Multi-flavour mass and decay constant identities}

For all $n$ we are using the same $O(p^4)$ effective Lagrangian
as for the $n=1$ case of Ref.~\cite{GL2} (except for one additional 
invariant 
$\langle\nabla^\mu U^\dag \nabla^\nu U\nabla_\mu U^\dag\nabla_\nu U\rangle$,
irrelevant for our purpose). The LEC's have an a priori unknown
dependence on $n$, therefore we  keep the same notation for all LEC's
with an extra index $n$. This is justified to the extent that the 
symmetry properties are as described above.  

The chiral expansion of Goldstone boson masses is very close to the case
$n=1$ discussed in Ref.~\cite{D}:
\begin{eqnarray}\label{pinmass}
F_\pi^2 M_\pi^2 &=& 2m\Sigma(n+2)+ 2m(n m_s +2m)Z^s_n+4m^2 A_n \nonumber \\
                & & +4 m^2 B_{0n}^2 L + F_\pi^2 M_\pi^2 d_{\pi n}\, ,    \\
F_K^2M_K^2      &=& (m_s + m)\Sigma (n+2) +(m_s+m)(nm_s+2m) Z^s_n \nonumber \\
                & & + (m_s+m)^2 A_n  + m(m_s + m)B_{0n}^2 L  \nonumber \\
                & & + m_s(m_s+m)B_{0n}^2 L_n'  + F_K^2 M_K^2 d_{K n}\, ,  \label{kanmass}\\
F_X^2M_X^2      &=& 2m_s \Sigma (n+2) + 2m_s(nm_s+2m)Z^s_n + 4m_s^2 A_n
\nonumber \\ 
                & & + 4m_s^2 B_{0n}^2  L'_n + F_X^2 M_X^2 d_{X n}\,,   \label{xnmass}
\end{eqnarray}
and likewise for $F_\eta^2 M_\eta^2$ (see App.~\ref{app:etax}). 
The connection with the standard LEC's of the $N_f \ge 3$ effective Lagrangian is:
\begin{eqnarray}
Z^s_n &=&  32 B_{0n}^2  \Biggl\{ L^n_6(\mu)  \nonumber \\
      & & -\frac{1}{512\pi^2}\left[\log\frac{M_K^2}{\mu^2} 
          + \frac{2}{(n+2)^2}\log \frac{M_\eta^2}{\mu^2} \right] \Biggr\}
       \,,\label{nZ} \\
A_n   &=&  16 B_{0n}^2  \Biggl\{ L^n_8(\mu) \nonumber \\  
      & & - \frac{n}{512\pi^2}\left[ \log\frac{M_K^2}{\mu^2}
      +\frac{2}{n+2} \log 
\frac{M_\eta^2}{\mu^2} \right] \Biggr\}\,. \label{nA}
\end{eqnarray}
We have the ratio:
\begin{equation}\label{nB}
B_{0n} = \frac{\Sigma(n+2)}{F^2(n+2)}\,,
\end{equation}
and $\Sigma(n+2)$ and $F(n+2)$ denote the condensate and the pseudoscalar
decay constant
for all $n+2$ massless quarks. The remainders $F_P^2M_P^2 d_P^2$ collect all
higher-order terms, starting at the next-to-next-to-leading order $O(m_q^3)$ (NNLO)
[hence $d_n=O(m_q^2)$]. The ($n$-independent) combination of chiral 
logarithms $L$ has the same meaning as in Ref.~\cite{D}:
\beq
L=\frac{1}{32\pi^2}
 \left[3\log\frac{M_K^2}{M_\pi^2}+\log\frac{M_\eta^2}{M_K^2}\right].
\eeq
whereas $L'$ contains all chiral logarithms involving the unphysical mass $M_X$:
\begin{equation}\label{Lprime}
 L'_n = \frac{n-1}{16\pi^2 n}\left\{\log\frac{M_\eta^2}{M_K^2}
                  - (n+1) \log\frac{M_X^2}{M_K^2}\right\}\,,
\end{equation}
which vanishes for $n=1$ as it should. 
Similar expressions are derived for the decay constants:
\begin{eqnarray}
F_\pi^2 &=& F^2(n+2) + 2(nm_s+2m) \tilde\xi_n + 2m \xi_n  \label{pindecay}\\
        & & +\frac{2mB_{0n}}{16\pi^2}  \left\{\log\frac{M_\eta^2}{M_K^2}
          + 2\log\frac{M_K^2}{M_\pi^2}\right\} +F_\pi^2 e_{\pi n}\,,
           \nonumber\\
F_K^2   &=& F^2(n+2) + 2 (nm_s + 2m) \tilde\xi_n + (m_s+m)\xi_n \label{kandecay}\\
        & & \quad + m B_{0n} L + m_s B_{0n} L'_n  +F_K^2 e_{K n}\,,\nonumber \\
F_X^2   &=& F^2(n+2) + 2 (nm_s + 2m)\tilde\xi_n +2m_s\xi_n  \label{xndecay}\\
        & & +\frac{2m_sB_{0n}}{16\pi^2}\left\{ \log\frac{M_\eta^2}{M_K^2}-
             n\log\frac{M_X^2}{M_K^2} \right\}+ F_X^2 e_{X n}\,.
           \nonumber
\end{eqnarray}
while $F_\eta$ is given in App.~\ref{app:etax}.
The scale-invariant constants $\xi_n$ and $\tilde\xi_n$ are related to
the LEC's $L_4$ and $L_5$ as follows:
\begin{eqnarray}\label{nxitilde}
\tilde \xi_n &=& 8B_{0n}\left\{ L^n_4(\mu) 
   -\frac{1}{256\pi^2}\log\frac{M_K^2}{\mu^2}\right\}\,,\\
\xi_n &=& 8B_{0n}\Biggl\{L^n_5(\mu) \nonumber \\
      & & - \frac{1}{256\pi^2}
    \left[n\log\frac{M_K^2}{\mu^2}
              + 2 \log\frac{M_\eta^2}{\mu^2}\right]\Biggr\} \,.
 \label{nxi}
\end{eqnarray}
The remainders $F_P^2 e_{P n}$ collect the higher-order terms, starting
at NNLO $[O(m_q^2)]$. 

\subsection{Low-energy constants} \label{sec:lecs}

It is a simple exercise of algebra to combine 
Eqs.~(\ref{pinmass})-(\ref{kanmass}),
and Eqs.~(\ref{pindecay})-(\ref{kandecay}), to
get the expression of the LEC's: 
\begin{eqnarray}\label{L6}
& & \frac{2m}{F_{\pi}^2 M_{\pi}^2} [\Sigma(n+2) + (nr+2)mZ^S_n] \nonumber \\
&=& 1-
 \epsilon(r)- d_n - \frac{4m^2 B_{0n}^2}{F_{\pi}^2 M_{\pi}^2}
 \frac{r}{r-1}(L-L'_n)\,,\\
\label{L8}
\frac{4m^2A_n}{F_{\pi}^2 M_{\pi}^2} &=& \epsilon(r) + d'_n     
+\frac{4m^2B_{0n}^2}{F_{\pi}^2 M_{\pi}^2} \frac{1}{r-1}(L-rL'_n)\,,
\end{eqnarray}
with 
\begin{equation}
\epsilon(r) = 2\frac{r_2-r}{r^2-1}, \quad
r_2= 2\left(\frac{F_KM_K}{F_{\pi}M_{\pi}}\right)^2 -1\,.
\end{equation}
$d_n$ and $d'_n$ are linear combinations of the remainders $d_{\pi n}$
and $d_{K n}$:
\begin{equation} \label{remainmass}
d_n = \frac{r+1}{r-1}d_{\pi n} - 
   \left(\epsilon(r)+\frac{2}{r-1}\right)d_{Kn}\,,\quad
d'_n=d_n-d_{\pi n}\,.
\end{equation}
A similar work can be performed for the decay constants:
\begin{eqnarray}
\frac{2m\xi}{F_\pi^2} 
&=&\eta(r)+e'_n  -\frac{2mB_{0n}}{(r-1)F_\pi^2}\Biggl[L+r L'_n \nonumber \\
& &    -\frac{1}{8\pi^2}\left(\log\frac{M_\eta^2}{M_K^2}
        +2\log\frac{M_K^2}{M_\pi^2}\right)\Biggr]\,,\label{L5} \\
(nr+2)\frac{2m\tilde\xi}{F_\pi^2} &=&
1-\eta(r)-e_n-\frac{F^2(n+2)}{F_\pi^2} \nonumber \\
& & +\frac{2mB_{0n}}{(r-1)F_\pi^2}  \Biggl[L +r L'_n \nonumber \\
& &  -\frac{r+1}{16\pi^2}\left(\log\frac{M_\eta^2}{M_K^2}
        +2\log\frac{M_K^2}{M_\pi^2}\right)
    \Biggr]\,,  \label{L4}
\end{eqnarray}
with:
\begin{eqnarray}
\eta(r)&=&\frac{2}{r-1}\left(\frac{F_K^2}{F_\pi^2}-1\right)\,,\\
e_n&=&\frac{r+1}{r-1}e_{\pi n}-
 \left(\eta(r)+\frac{2}{r-1}\right)e_{K n}\equiv e'_n+e_{\pi n}
 \,. \label{remaindecay}
\end{eqnarray}
We recall that $\epsilon(r)$ and $\eta(r)$ are suppressed at large
values of $r$ ($\geq 15$). We expect then $d'_n \ll d_n \sim d_{\pi n}$
and $e'_n\ll e_n\sim e_{\pi n}$.

One easily checks that all the formulae displayed in this paragraph 
reduce to the mass and decay constant identities 
obtained in Ref.~\cite{D} in the case $n=1$.

\subsection{$SU(2)\times SU(2)$ order parameters}

                In the $SU(n+2)\times SU(n+2)$ chiral limit
($m=m_s=0$), we define the dimensionless order parameters:
\begin{eqnarray}\label{order-3-flav}
X(n+2) &=& \frac{2m \Sigma(n+2)}{F^2_\pi M^2_\pi}\,, \nonumber\\
Y(n+2) &=& \frac{2m B_{0n}}{ M^2_\pi}\,, \\
Z(n+2) &=& \frac{F^2(n+2)}{F^2_\pi}\,. \nonumber
\end{eqnarray}
The Gell-Mann--Oakes--Renner ratio $X(n+2)$ measures the quark condensate in
physical units, while $Z(n+2)$ does the same for the decay constant.
We have $X(n+2)=Y(n+2)Z(n+2)$.

We consider now the $SU(2)\times SU(2)$ chiral limit where only $m$ vanishes (and
the $n$ copies of the strange quark remain massive), in order to investigate the
effect of massive sea quarks on two-flavour chiral dynamics. We define the quark
condensate and the decay constant in this limit as:
\begin{equation}
\Sigma(2)=\lim_{m\to 0} \frac{F_\pi^2M_\pi^2}{2m}\,,\quad
F^2(2)=\lim_{m\to 0} F^2_\pi\,,
\end{equation}
and the dimensionless order parameters:
\begin{eqnarray}\label{order}
X(2) &=& \frac{2m \Sigma(2)}{F^2_\pi M^2_\pi}\,, \nonumber \\ 
Z(2) &=& \frac{F^2(2)}{F^2_\pi}\,,\\ 
Y(2) &=& \frac{X(2)}{Z(2)} = \frac{2m B(2)}{M^2_\pi}\, . \nonumber
\end{eqnarray}

From Eqs.~(\ref{pinmass}) and (\ref{L6}), we derive the expression for the
two-flavour condensate:
\begin{eqnarray} 
X(2)[1-\bar{d}_{\pi n}] &=& \frac{nr}{nr+2}
\biggl[1-\epsilon(r) -d_n-[Y(n+2)]^2 f_n \biggr]  \nonumber \\
& & +\frac{2}{nr+2} X(n+2)\,, \label{xtwo}
\end{eqnarray}
where a quantity topped with a bar is considered in the limit $m\to 0$ and:
\begin{eqnarray}
f_n &=& \frac{M_\pi^2}{F_\pi^2}\left(\frac{r}{r-1} [L-L'_n] - \frac{nr+2}{2}
\Delta Z^s_n\right)\,,  \label{deffn}\\
\bar{Z}^s_n &=& Z^s_n+B_{0n}^2\Delta Z^s_n\,, \nonumber \\
\Delta Z^s_n &=& \frac{1}{16\pi^2}\left[\log\frac{M_K^2}{\bar{M}_K^2} 
       + \frac{2}{(n+2)^2}\log \frac{M_\eta^2}{\bar{M}_\eta^2} \right]\,. \label{deltazs}
\end{eqnarray}
In a similar fashion, the two-flavour decay constant can be read from
Eqs.~(\ref{pindecay}) and (\ref{L4}):
\begin{eqnarray} 
Z(2)[1-\bar{e}_{\pi n}] &=& \frac{nr}{nr+2}
\left[1-\eta(r)-e_n-Y(n+2)g_n\right] \nonumber \\
&& +\frac{2}{nr+2} Z(n+2)\,, \label{ztwo}
\end{eqnarray}
with the quantities:
\begin{eqnarray}
g_n &=& \frac{M_\pi^2}{F_\pi^2}\Biggl(\frac{r}{r-1} [L-L'_n]
   + \frac{r+1}{r-1} \frac{1}{32\pi^2} \log\frac{M_\eta^2}{M_\pi^2}
\nonumber \\
    & &   - (nr+2) \Delta\tilde\xi^s_n\Biggr)\,. \label{defgn}\\
\bar{\tilde\xi}_n&=&\tilde\xi_n+B_{0n}\Delta \tilde\xi_n\,, \quad
\Delta \tilde\xi_n=\frac{1}{32\pi^2}\log\frac{M_K^2}{\bar{M}_K^2}\,. \label{deltaxit}
\end{eqnarray}
$f_n$ and $g_n$ contain chiral logarithms of two different types and signs:
the first involve the masses of the Goldstone
bosons $M_{\pi,K,\eta,X}$ and are positive, while the latter 
are negative and combine ratios of masses of the same meson, considered in the
massive theory and in the $SU(2)\times SU(2)$ chiral limit ($m\to 0$).
We can estimate $f_n$ and $g_n$ by iterating the previous mass and
decay constant identities as explained in Apps.~\ref{app:etax} and \ref{app:fg}.

\section{Connection between fluctuation and order parameters} \label{sec:connect}
\label{sec:4}
                     We want now to investigate the connection between     
chiral order parameters and vacuum fluctuations. 
For a generic $n>0$, the mass and decay
constant identities yield a very simple expression of the two order
parameters of fundamental interest, namely the condensate $\Sigma(n+2)$ and
the decay constant $F(n+2)$, in terms of the two OZI-rule violating LEC's
$L_4$ and $L_6$. These two large-$N_c$ suppressed constants directly
reflect vacuum fluctuations of $\bar qq$ pairs and they provide a convenient
framework to discuss completely and transparently how these fluctuations
affect order parameters.

The first step consists in eliminating the LEC's $A_n$ and $\xi_n$
from the mass and decay constant identities for the pion and the kaon
Eqs.~(\ref{pinmass})-(\ref{kanmass}) and (\ref{pindecay})-(\ref{kandecay}),
leading to Eqs.~(\ref{L6}) and (\ref{L4}).
Then we can reexpress the (OZI-rule violating)
constants $\tilde \xi_n$ and $Z^s_n$ in terms of $L_4$ and $L_6$, and
obtain the two desired relations between the order parameters $X,Y,Z$ and
the fluctuation parameters $L_4$ and $L_6$:
\begin{eqnarray}
X(n+2)&=&1-\epsilon(r)-d_n-[Y(n+2)]^2\rho_n/4\,, \label{xl6}\\
Z(n+2)&=&1-\eta(r)-e_n-Y(n+2)\lambda_n/4\,.      \label{zl4}
\end{eqnarray}

The constants $L_6$ and $L_4$ enter the discussion through the combinations:
\begin{equation}\label{lambdarho}
\rho_n = 64 \frac{M_\pi^2}{F_\pi^2} (nr + 2) \Delta L_6^n\,, \quad
\lambda_n = 32 \frac{M_\pi^2}{F_\pi^2} (nr + 2) \Delta L_4^n\,,
\end{equation}
where the scale-independent differences
$\Delta L_i^n=L_i^n(\mu)-L_i^{n,{\rm crit}}(\mu)$ involve the critical
values of the LEC's defined as:
\begin{eqnarray}
L_4^{n,{\rm crit}}(\mu)
          &=& \frac{1}{256\pi^2}
          \log\frac{M_K^2}{\mu^2}-\frac{r}{8(r-1)(nr+2)}[L-L'_n]\nonumber\\
 &&\quad -\frac{1}{256\pi^2}\frac{r+1}{(r-1)(nr+2)}
           \log\frac{M_\eta^2}{M_\pi^2}
         \label{ncritical4} \,,\\
L_6^{n,{\rm crit}}(\mu) &=& \frac{1}{512\pi^2}\left(
       \log\frac{M_K^2}{\mu^2} 
       + \frac{2}{(n+2)^2}\log \frac{M_\eta^2}{\mu^2}
         \right)\nonumber\\
   &&\quad -\frac{r}{16(nr+2)(r-1)}[L-L'_n]\,.\label{ncritical6}
\end{eqnarray}

        Comparing the fluctuation parameters $\lambda_n$ and $\rho_n$ 
(\ref{lambdarho}) to 1 provides a quantitative measure of the effect 
of the LEC's $L_4$ and $L_6$
on observable quantities. The effect disappears if $L_4$ and $L_6$ are fine-tuned
to their critical values (\ref{ncritical4}) and (\ref{ncritical6}),
which for $n=1$, $r=25$, $\mu= M_\rho$
become 
$L_4^{\rm crit}(M_\rho)=-0.51 \cdot 10^{-3}$ and $L_6^{\rm crit}(M_\rho)=-0.26\cdot 10^{-3}$.
 
Notice that even a small deviation from the
critical values is amplified by the large numerical coefficients
in Eq.~(\ref{lambdarho}). 
It is possible and convenient to absorb the NNLO and higher-order 
contributions -- represented in Eqs.~(\ref{xl6})-(\ref{zl4})
by the remainders $d_n$ and $e_n$ --
into a multiplicative renormalisation of order and fluctuation parameters.
Defining the rescaled parameters as:
\begin{eqnarray}\label{yzrescaled}
z_n = \frac{F^2(n+2)}{F_\pi^2 [1-\eta(r) -e_n]} \,, &\quad&
y_n = \frac{1- \eta(r)-e_n}{1- \epsilon(r)- d_n} \frac{2mB_{0n}}{M_\pi^2}\,,
\nonumber \\
&&\\
\label{uvrescaled}
u_n = \lambda_n k_n, \quad  v_n = \rho_n k_n,
&\quad&
k_n(r) =\frac{1-\epsilon(r)-d_n}{(1-\eta(r)-e_n)^2}, \nonumber \\
&& 
\end{eqnarray}
the two basic equations inferred from the mass and decay constant identities 
take now the concise form:
\begin{eqnarray}\label{u}
z_n + \frac{1}{4} u_n y_n &=& 1\,,\\ 
\label{v}
z_n +\frac{1}{4} v_n y_n &=& \frac{1}{y_n}\,.
\end{eqnarray}

In addition, the multi-flavour quark condensate [expressed in GOR units,
cf. Eq.~(\ref{order-3-flav})] can be rewritten in terms of $y_n$ and $z_n$:
\beq \label{Xyz}
X(n+2) = [1-\epsilon(r)-d_n] y_n z_n\,.
\eeq

        At this place, a few remarks are in order:
        
\emph{i)} The above analysis holds for a generic $n>0$, including the physical case
$n=1$ of three flavours. On the other hand, the case of two massless flavours 
($m = 0$, $m_s$ nonzero) requires a separate discussion based on Eqs.~(\ref{xtwo}) and
(\ref{ztwo}) above.

\emph{ii)} The relations (\ref{u}) and (\ref{v}) between the (rescaled) order parameters
$y,z$ on one hand and the fluctuation parameters $u,v$ on the other hand
represent exact identities which do not result from any approximation or
expansion. The influence of higher chiral orders -- NNLO and beyond --
is entirely encoded in the rescaling factors (\ref{yzrescaled}) 
and (\ref{uvrescaled}) through the remainders $d_n$ and $e_n$.  
The latter, defined in Eqs.~(\ref{remainmass}) and (\ref{remaindecay}), stem from
$O(p^6)$ (and higher) contributions to the mass and decay constant identities.
They are both of order $d_n,e_n = O(m_q^2)$ and should remain small
unless the whole $\chi$PT series blows up.

\emph{iii)} We expect the rescaling factors in Eqs.~(\ref{yzrescaled}) and 
(\ref{uvrescaled}) to be close to 1. Two circumstances could spoil this expectation. 
First, if the quark mass ratio $r$ were small (typically $r< 10$), 
the quantity $\epsilon(r)$ would get close to 1.
On the other hand, for $r>15$ one has $\epsilon(r)< 0.2$ and $\eta(r) < 0.07$.
Such higher values of $r$ are preferred by the new low-energy $\pi\pi$ scattering
data \cite{DFGS}, discussed in Sec.~\ref{sec:twoflav}. 
Second, even if the remainders $d_n$ and $e_n$ are small
in the physical case $n = 1$ (say 10 \% or less), their
size could grow when $n$ increases. We shall come back to the 
large-$n$ behaviour of the theory in Sec.~\ref{sec:largefluc}.
In either case, we would not be allowed to treat perturbatively the 
rescaling factors as close to 1.

\emph{iv)} Let us consider a solution of the generic system (\ref{u}) and (\ref{v}) giving
the order parameters $y$ and $z$ in terms of the fluctuation parameters $u$ and $v$.
We can then read from Eqs.~(\ref{L8}) and (\ref{L5}) 
together with Eqs.~(\ref{nA}) and (\ref{nxi}) the
following parameters of $\leff$: $2mB_{0n}$, $F_0^2 = F^2(n+2)$, $L_5(\mu)$ and
$L_8(\mu)$~\footnote{
If we include the identities for $\eta$, 
we can also express the constant $L_7$ -- see App.~\ref{app:etax}.}
as functions of $r=m_s/m$, $L_4(\mu)$, $L_6(\mu)$ the NNLO remainders
$d_{\pi n}$, $d_{K n}$ and $e_{\pi n}$, $e_{K n}$~\footnote{
Alternatively, we can express all $O(p^4)$ symmetry breaking LEC's as 
functions of $r$ and the two order parameters $X(n+2)$ and $Z(n+2)$,
as discussed in Ref.~\cite{D} for $n=1$.}.
These expressions are then used in the study of other observables (e.g. Goldstone Boson
scattering amplitudes) to eliminate $O(p^2)$ and $O(p^4)$ LEC's from the bare
$\chi$PT formulae. In particular, this procedure should be used when the NNLO
remainders are (iteratively) matched with two-loop $\chi$PT expressions. 
The procedure described above -- eliminating constants of $\leff$
in favour of observables -- differs from the one used in standard $\chi$PT, 
unless the fluctuation parameters $u$ and $v$ are small compared to 1. This
might in particular affect the outcome of standard analyses beyond
one loop~\cite{ABT}.

\emph{v)}       If $u_n,v_n\ll 1$, the standard multi-flavour $\chi$PT may be recovered
in two steps. First, one constructs the perturbative solution
of the non-linear system (\ref{u}) and (\ref{v}) in powers of 
$u_n , v_n = O(m_q)$:

\begin{eqnarray}
y_n &=& 1 + \frac{1}{4} (u_n - v_n)  + \frac{1}{8} (u_n - v_n )^2 + \ldots \label{perty}\\
z_n &=& 1 - \frac{1}{4} u_n - \frac{1}{16} u_n ( u_n - v_n ) +  \ldots     \label{pertz}
\end{eqnarray}

Next, one returns to the original variables $\lambda_n$, $\rho_n$ in Eq.~(\ref{lambdarho}) and
to $2m B_{0n} / M_\pi^2$, $F^2(n+2) / F_\pi^2$ expanding the rescaling factors in
Eqs.~(\ref{yzrescaled}) and (\ref{uvrescaled}) around 1 in powers of $\epsilon(r) = O(m_q)$, 
$\eta(r) = O(m_q)$ and of the remainders $d_n, e_n = O(m_q^2)$. Matching the
latter with the explicit two-loop contributions 
to Eqs.~(\ref{pinmass})-(\ref{kanmass}) and
(\ref{pindecay})-(\ref{kandecay}), one reproduces the standard $\chi$PT expansion up to and
including $O(p^6)$ order.

\emph{vi)} In the physical situation of three massless flavours ($n=1$), the fluctuation
parameters need not be small enough to allow the power expansion of
Eqs.~(\ref{perty})-(\ref{pertz}):
using the estimates for $L_6^1 (M_\rho) = 0.6 \cdot 10^{-3}$ (central value)
based on sum rules for the correlator
$\langle(\bar{u}u + \bar{d}d) \bar{s}s\rangle $~\cite{Bachir1,Bachir2,D} and
the recent determination of $L_4^1(M_\rho) = 0.1\cdot 10^{-3}$ (central value)
from $\pi K$ scattering data~\cite{Kpi}, one obtains the rough estimate $u_1 \sim 1.2$,
$v_1 \sim 3.4$ (for $r=25$). 
As pointed out in Ref.~\cite{DGS}, important vacuum fluctuations of
$\bar{q}q$-pairs suppress the three-flavour condensate and destabilise the $\chi$PT
expansion. 

This phenomenon is a particular consequence of Eqs. (\ref{u}) and (\ref{v}).
Multiplying the latter by $y$ and using Eq. (\ref{Xyz}), one obtains the
relation: 
\beq \label{xresum}
X(n+2) = 2 \frac{1 - \epsilon(r) - d_n}{1 + \sqrt{1 + v_n/z_n^2}}\,,
\eeq
which is identical (in the physical case $n=1$) to Eq.~(9) of Ref.~\cite{DS}.

To cope with possibly large values of the fluctuation parameters $u$ and $v$,
it may very well be sufficient to replace the perturbative solution (\ref{perty})-(\ref{pertz})
of the system (\ref{u})-(\ref{v}) by its exact algebraic solution,
and to keep the perturbative expansion of the rescaling 
factors in Eqs.~(\ref{yzrescaled}) and (\ref{uvrescaled}).
Let us emphasise that the divergence of the power series 
(\ref{perty})-(\ref{pertz}) is a question logically disconnected from 
the convergence of the expansion of rescaling factors around 1: 
the former is related to a possible suppression of
multi-flavour condensate and of the corresponding leading order of 
$\chi$PT, whereas the latter is more a question of a global 
convergence of the $\chi$PT series starting at NNLO order.

            In the remaining sections of this paper we concentrate on the
non-perturbative analysis of the system (\ref{u})-(\ref{v}) and its consequences
for the breaking of both $SU(n+2) \times SU(n+2)$ and $SU(2) \times SU(2)$ chiral
symmetries in various limits.

\section{Positivity and paramagnetic constraints}
\label{sec:5}

In this section we investigate the system of equations (\ref{u})-(\ref{v}) in
the light of the positivity of the order parameters $X(n+2)$, $Y(n+2)$ and $Z(n+2)$
and of the conjectured paramagnetic inequalities they have to satisfy~\cite{DGS}.
We are mainly interested in the domain allowed by these constraints in
the plane of the fluctuation parameters $u$ and $v$.  For simplicity, we omit
further reference to $n$ and $r$, since the latter enter the system (\ref{u})-(\ref{v})
only via rescaling factors. The system admits two solutions
for the ratio of the order parameter $y$ (related to $B_{0n}$):
\begin{equation}\label{ysol}
y_+= \frac{2}{1+\sqrt{1+v-u}}\,, \quad \quad y_-=\frac{2}{1-\sqrt{1+v-u}}\,.
\end{equation}
These two solutions depend on the difference of the two
rescaled fluctuation parameters $u$ (function of the LEC $L_4$) and
$v$ (function of $L_6$). Notice that $v-u$ is related to the particular
combination $2 L_6 - L_4$. The square root in Eq.~(\ref{ysol}) signals a
non-perturbative resummation of vacuum fluctuations as discussed in the previous
section. For $|v-u|\ll 1$, the Taylor expansion of $y_+$ reproduces the $\chi$PT
series, whereas $y_-$ is of truly non-perturbative nature.

\begin{figure}[ht]
\epsfysize8cm
\centerline{\epsffile{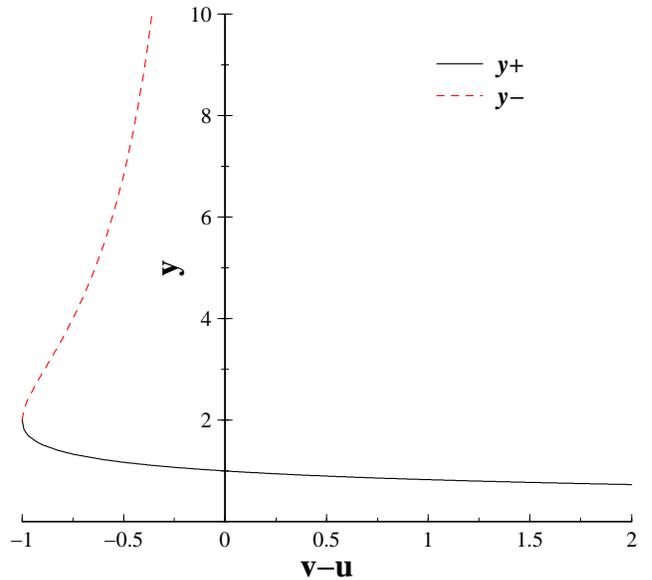}
}
\caption{The two branches for $y$ as functions of $v-u$.
The solid (black) branch is $y_+$, while the dashed (red)
branch is $y_-$.}
\label{fig:ybranches}
\end{figure}

         The two branches  $y_+$ and $y_-$ can be considered as two different
sheets of the two-valued function $y(v,u)$. These sheets are tangent
along the line $v - u = -1$ that coincides with the boundary of the  
definition domain $v>u-1$. Along this boundary, $y$ has the constant value 2.
The two branches $y_+$ and $y_-$ are drawn in Fig.~\ref{fig:ybranches} as functions of
$v-u$. When $v - u$ increases, $y_+$ decreases and vanishes asymptotically at
infinity, whereas $y_-$ increases and tends to infinity for  $v - u\to 0^-$.
         
        The system (\ref{u})-(\ref{v}) yields the ratio $y$ and the (rescaled)
decay constant $z$, both of which should be positive for an acceptable solution:
the vacuum stability requires $x=yz>0$ and $z$ is related to $F^2(n+2)/F^2_\pi$.
As far as $y$ is concerned, $y_+$ is positive in the whole half-plane $v - u > -1$ (where
$0< y < 2$), and $y_-$ is positive inside the strip $-1 < v - u < 0$ (where $y > 2$).
The positivity of $z$, i.e. the condition  $z = 1 - u y/4 >0$, yields additional
constraints in the $(u,v)$ plane, especially for $u>0$. The critical line
$z = 0$ is the parabola $v = u^2 / 4$, along which $y = 4 / u$. The condition
$z > 0$ is trivially satisfied for negative  $u$, but for $u > 0$ it leads
on both sheets to (different) non-trivial bounds. 

\begin{figure}
\centerline{
\epsfysize5cm\epsffile{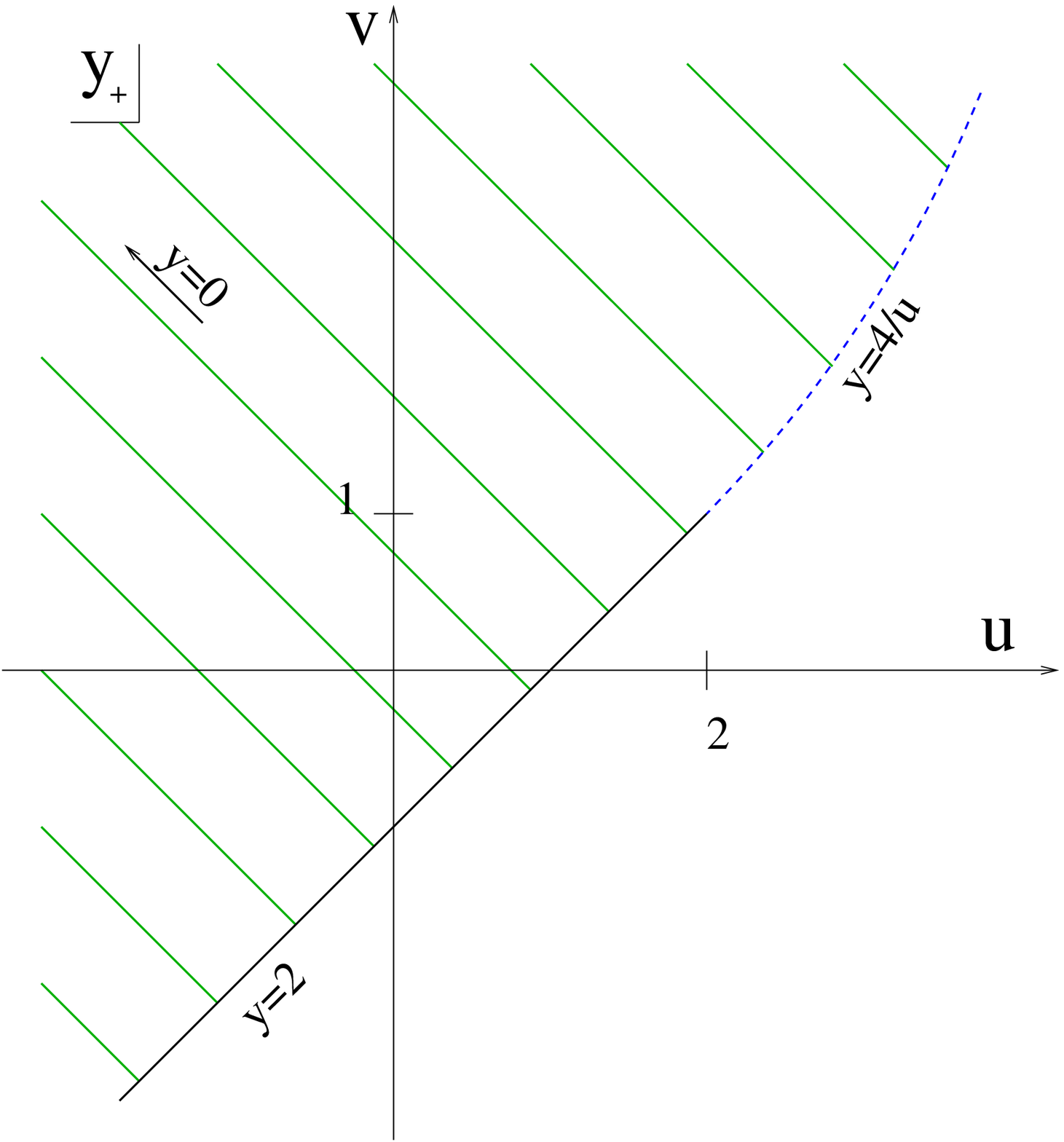}\quad
\epsfysize5cm\epsffile{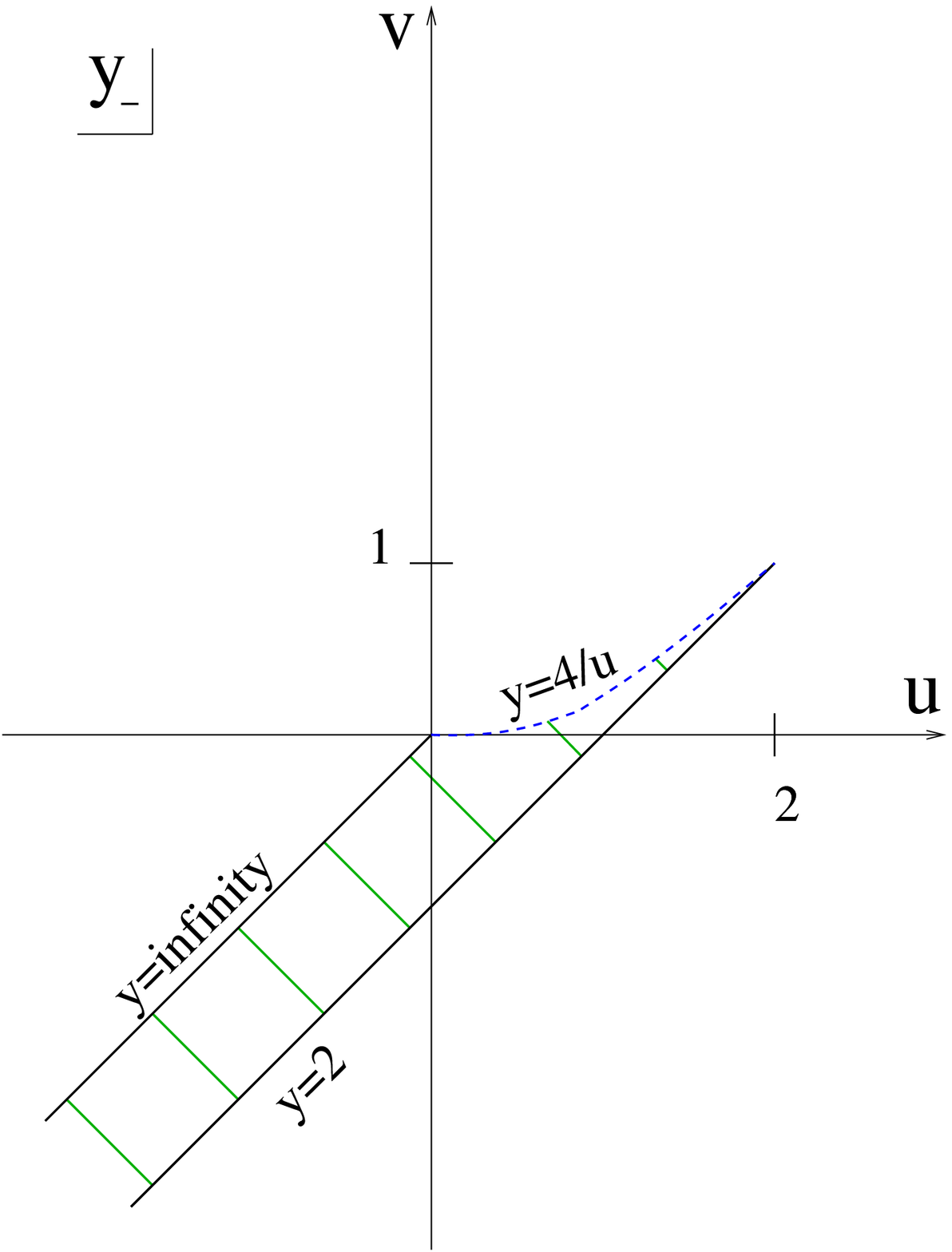}
}
\caption{The two sheets for $y$ in the $(u,v)$ plane: on the
left, the $y_+$ sheet, on the right, the $y_-$ sheet. The hatched (green) 
regions are the allowed domains where $y$ and $z$ are both positive.
The value of $y$ at each boundary is indicated. The critical line
$z=0$ of symmetry restoration is denoted with a dashed (blue) line.}
\label{fig:twosheets}
\end{figure}

  As a result the whole domain of positivity of both $y$ and $z$ is obtained. 
On the $y_+$ sheet, one must have $v> u-1$ for $u<2$, and $v > u^2 / 4$ for $u>2$.
On the $y_-$ sheet, the positivity domain amounts to the part of the strip
$u > v > u- 1$ situated below the parabola $v = u^2 / 4$. These two domains
are represented in Fig.~\ref{fig:twosheets}.

We now arrive at the constraints imposed by the paramagnetic
inequalities between order parameters discussed in Ref.~\cite{DGS}. It was suggested 
that the chiral order parameters that are dominated by the lowest
eigenvalues of the Euclidean Dirac operator are particularly sensitive
to a paramagnetic suppression arising from the infrared part of the fermion
determinant -- i.e. from the light-quark loops. This applies in particular to
the decay constant $F^2$ and to the quark condensate $\Sigma$: 
the more flavours from the total of $N_f = n + 2$ become massless, 
the more suppressed the fundamental order parameters get.
As a corollary, $\Sigma (2)$ should be an increasing
function  of the strange quark mass $m_s$, as long as the latter is comparable
to the QCD scale. This suggests the two paramagnetic constraints:
\begin{equation} \label{param}
X(2)\geq X(n+2)\,, \quad \quad Z(2) \geq Z(n+2)\,,
\end{equation}
which can be reexpressed using the expressions of $X(2)$ in Eq.~(\ref{xtwo})
and $Z(2)$ in Eq.~(\ref{ztwo}). They can be further simplified using the notation
of the previous section, and yield for $X$ and $Z$ respectively:
\begin{equation} \label{paramuv}
X:v \geq  \frac{4}{1-D} \left[k f - \frac{D}{y^2}\right]\,,\quad
Z:u \geq  \frac{4}{1-E} \left[k g - \frac{E}{y}\right]\,,
\end{equation}
where the NNLO remainders are rescaled:
$D_n = \bar{d}_{\pi n} \cdot (nr+2)/(nr)$,
$E_n = \bar{e}_{\pi n} \cdot (nr+2)/(nr)$,
and $f$ and $g$ are positive combinations of chiral logarithms
defined in Eqs.~(\ref{deffn}) and (\ref{defgn}) and estimated in App.~\ref{app:fg}.

The paramagnetic inequalities Eq.~(\ref{param}) yield therefore lower bounds for
the fluctuation parameters $u$ and $v$. These bounds are only useful if
the ratio of order parameters $y$ is large enough, for instance on the second
sheet where $y$ could grow. But we do not expect
Eq.~(\ref{paramuv}) to have much relevance, say, on the first sheet for large $u$ and $v$.
Using the system of Eqs.~(\ref{u})-(\ref{v}), we may convert these lower bounds on the 
fluctuation parameters into upper bounds for $y$. The constraints on
$X$ and $Z$ yield respectively:
\begin{equation}
\label{paramy}
\begin{array}{l}
\displaystyle 
X:y\leq \frac{2}{(1-D)z+\sqrt{(1-D)^2z^2+4 k f}}\,,\\
\displaystyle
Z:y\leq \frac{1-z(1-E)}{kg}\,.
\end{array}
\end{equation}

From Eq.~(\ref{paramy}) and the discussion of the previous section,
we obtain the important result that $y$ is necessarily non-vanishing and finite
on the physical domain of the two sheets:
\begin{equation} \label{boundsy}
0<y\leq \min\left(\frac{1}{\sqrt{k f}},\frac{1}{kg}\right)\,.
\end{equation}
Moreover, if $y$ is large enough, 
Eq.~(\ref{paramuv}) shows that $u$ is positive, and
thus $z=1-uy/4 < 1$. 
The paramagnetic inequalities in Eq.~(\ref{param}) lead therefore
to bounds for the two main $SU(n+2)\times SU(n+2)$ chiral order parameters $X$ and $Z$.

\begin{figure}[t]
\centerline{
\epsfysize7cm\rotatebox{270}{\epsffile{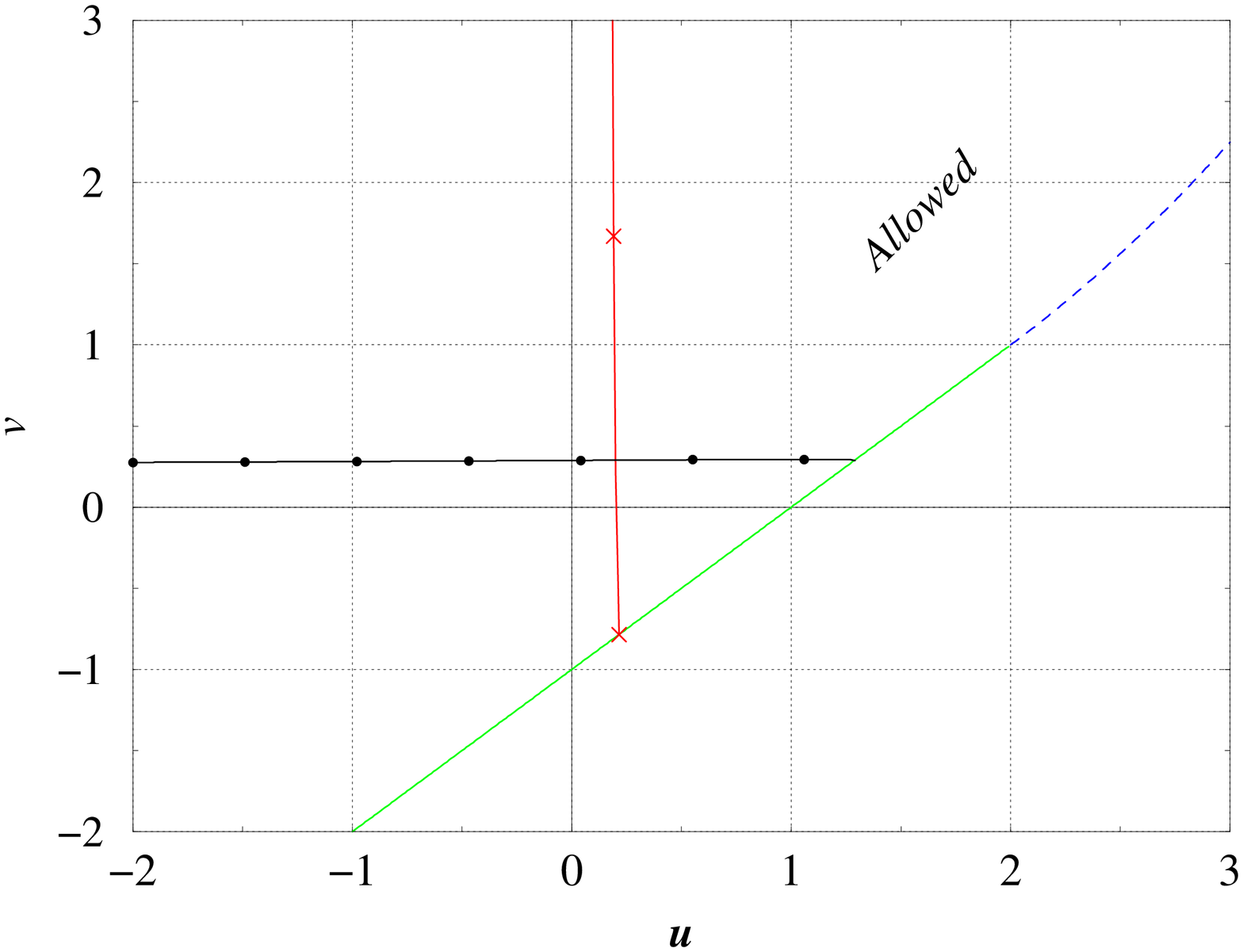}}}
\centerline{\epsfysize7cm\rotatebox{270}{\epsffile{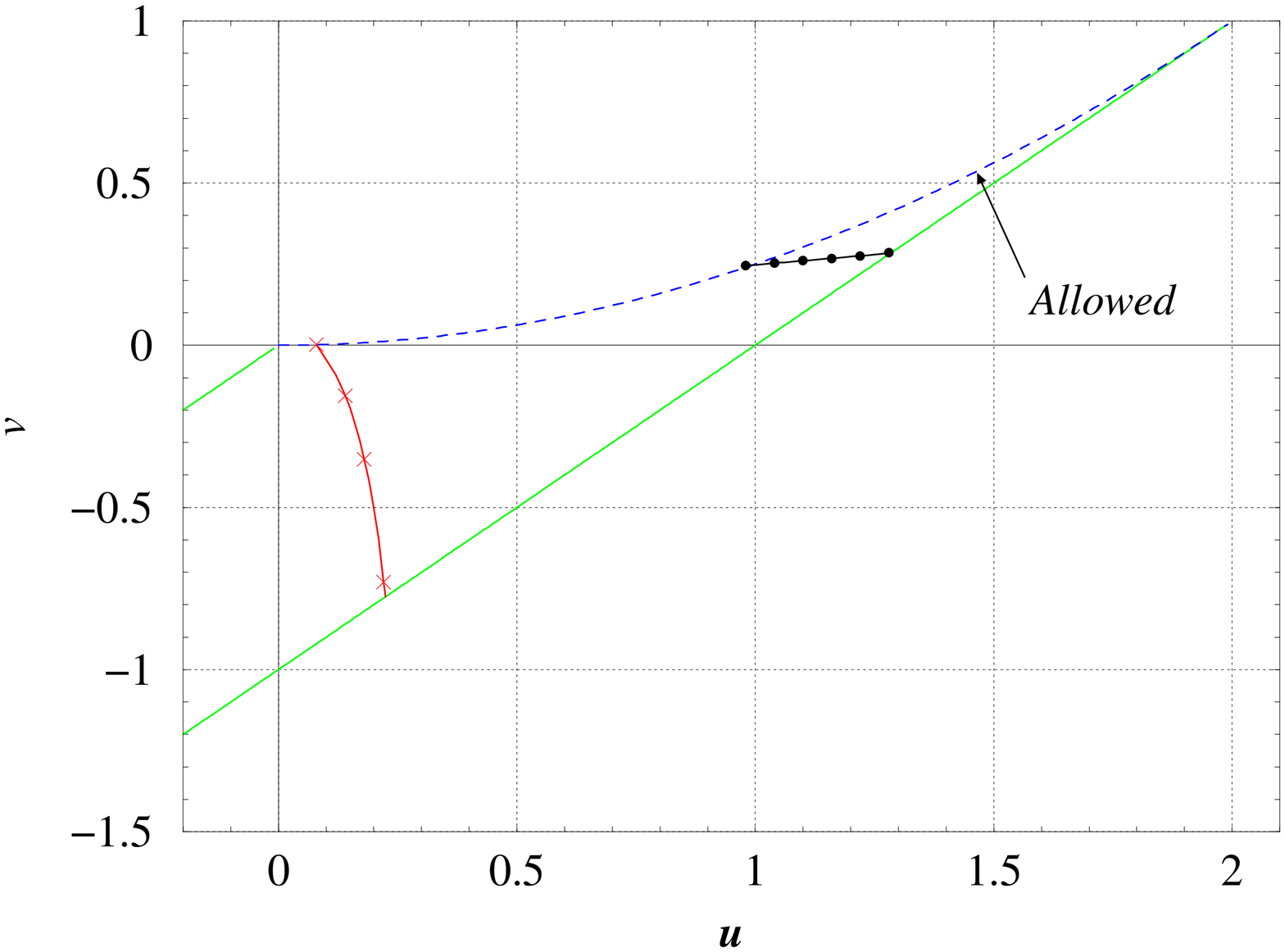}}
}
\caption{Illustration of the paramagnetic inequalities on the two sheets for 
$n=1$ and $r=25$ (upper: first sheet $y_+$, lower: second sheet $y_-$). 
The boundaries of the two sheets are indicated with solid (green) lines,
while the (blue) dashed line is the critical parabola $z=0$. The
(black) line with filled circles indicates the lower bound for $v$, derived from
the paramagnetic inequality for $X$. The
(red) line with crosses indicates the lower bound for $u$, derived from
the paramagnetic inequality for $Z$. The NNLO remainders $D$ and $E$ are set to 0.}
\label{fig:paramag}
\end{figure}

In the case $n=1$ and $r=25$, the estimation procedure detailed in 
App.~\ref{app:fg} leads to Fig.~\ref{fig:paramag}
for $u$ and $v$. We see that
only the upper right part of the $(u,v)$ plane survives. In particular, only a
small fraction of the second sheet, far from the origin, remains available. Let us
emphasise that the paramagnetic inequalities yield such constraints 
only if the two combinations of
chiral logarithms $f$ and $g$ are positive.

It is worth noting that the existence of minimal
values of fluctuation parameters $u$ and $v$ does not contradict
$\chi$PT which requires $u,v\to 0$ in the chiral limit. 
The minimal values (\ref{paramuv}) are indeed proportional to $M_\pi^2$
as the parameters $u$ and $v$ themselves, and therefore constrain only 
how quickly $u$ and $v$ vanish in the chiral limit.

As a conclusion, the system (\ref{u})-(\ref{v}) admits two solutions for the
ratio of order parameters $y$: they can be considered as the two sheets
of a two-valued function $y(v,u)$. In the $(u,v)$ plane of fluctuation parameters,
the domain of definition of the two sheets is restricted by 
positivity constraints [vacuum stability, $F^2(n+2)\geq 0$]
and by paramagnetic inequalities [$X(n+2)\leq X(2)$, $Z(n+2)\leq Z(2)$]. As a result,
the first sheet is limited to positive (and possibly very large) values of the 
fluctuation parameters $u,v$, whereas only a tiny region of the second sheet, far from the 
origin but below $u=2,v=1$, is allowed.

\section{The case of symmetry restoration}
\label{sec:6}

We investigate now the vicinity of the critical line where 
the $SU(n+2)\times SU(n+2)$ chiral symmetry  is restored.
Since we exploit results from the effective Lagrangian
describing the spontaneously broken phase in terms of Goldstone bosons, we 
can comment only on the approach to the chirally symmetric phase
from the broken one (and not on specific properties of the former phase, where the 
$SU(n+2)\times SU(n+2)$ chiral symmetry is restored).

The symmetry restoration is equivalent to the vanishing of the
 decay constant $F_\pi$ in the chiral limit, viz. $Z(n+2) = 0$. In the 
$(u,v)$ plane, the condition $z=0$ corresponds to the critical line
$v = u^2/4$. The limit $z\to 0^+$ can be reached on both sheets: on the $y_+$ sheet 
the critical line is approached from above and $u>2$, while on the $y_-$ sheet,
the parabola $v= u^2/4$ must be approached from below and $0<u<2$.
On both sheets, the ratio of order parameters has the value
$y=4/u$ on the critical line. 

We see that the vicinity of the origin in the $(u,v)$ plane on the second sheet
plays a special role in the discussion: $y$ diverges there. 
The paramagnetic inequalities Eq.~(\ref{boundsy}) however prevent us 
from reaching the vicinity of $u=v=0$ on the second sheet: 
$y$ is bounded and cannot become arbitrarily large. 
The ratio of order parameters $Y(n+2)$ remains thus finite 
when the critical line $Z(n+2)=0$ is approached, so that the quark 
condensate $X(n+2)=Y(n+2)Z(n+2)$ vanishes. This was expected since 
all $SU(n+2)\times SU(n+2)$ order parameters must vanish as
chiral symmetry is restored.

It is also interesting to study the order parameters 
defined in the $SU(2)\times SU(2)$ chiral limit $m=0$.
Eqs.~(\ref{ztwo}) and (\ref{xtwo}) yield the 
two-flavour decay constant and quark condensate:
\begin{eqnarray} 
Z(2)[1-\bar{e}_{\pi n}]
 &=&\frac{nr}{nr+2} [1-\eta(r)-e_n]\nonumber \\
 & &\times \left\{1-y_n k_n(r) g_n + \frac{2}{nr} z_n\right\} \label{z2simple} \\
&\stackrel{z_n\to 0}{\to}& 
   \frac{nr}{nr+2} [1-\eta(r)-e_n] \nonumber \\
   && \times \left\{1-\frac{4 k_n(r) g_n}{u_n}\right\}
   \label{z2crit}\,, \\
X(2)[1-\bar{d}_{\pi n}] 
 &=&\frac{nr}{nr+2} [1-\epsilon(r)-d_n] \nonumber \\
 & &\times \left\{1-y_n^2 k_n(r) f_n + \frac{2}{nr} y_n z_n\right\}
\label{x2simple} \\
 &\stackrel{z_n\to 0}{\to}&
     \frac{nr}{nr+2} [1-\epsilon(r)-d_n] \nonumber \\
     && \times \left\{1-\frac{16 k_n(r) f_n}{u_n^2}\right\}\,.   \label{x2crit}     
\end{eqnarray}

As long as we are not in the vicinity of the origin $u=v=0$,
we expect $X(2)$ and $Z(2)$ to remain close to 1 (for $r\geq 20$) along
the critical line where the $SU(n+2)\times SU(n+2)$ order parameters vanish.
We can understand it by combining the definition
of $X(2)$ and $Z(2)$, see Eq.~(\ref{order}), with the Ward identities for the
masses and decay constants:
\begin{eqnarray}
X(2)[1-\bar{d}_{\pi n}] &=& X(n+2) + n r \frac{2m^2 \bar{Z}^s_n}{F_\pi^2 M_\pi^2}\,,\\
Z(2)[1-\bar{e}_{\pi n}] &=& Z(n+2) + n r \frac{2m \bar{\tilde\xi}_n}{F_\pi^2}\,.
\end{eqnarray}
For instance, the two-flavour quark condensate is the sum of its
$SU(n+2)\times SU(n+2)$ counterpart and a LEC describing the violation of the OZI
rule in the scalar sector. We shall call the first term the ``genuine'' condensate
-- stemming directly from the breakdown of $SU(n+2)\times SU(n+2)$ chiral symmetry -- and
the latter the ``induced'' condensate -- induced by the massive 
strange-like quark pairs present in the $SU(2)\times SU(2)$ vacuum~\cite{GST}. 
The same analysis
applies to the decay constants $Z(2)$ and $Z(n+2)$.
We see now clearly how $SU(2)\times SU(2)$ chiral symmetry can remain broken while
the $SU(n+2)$ critical line is reached. Even though there is no genuine 
contribution to the two-flavour order parameters, the vacuum is 
not invariant under $SU(2)\times SU(2)$ chiral rotations because of
the symmetry-breaking transitions $\bar s_is_i \to \bar uu + \bar dd$ which
violate the OZI rule and are suppressed for $N_c\to\infty$.

Let us now move along the critical line towards the origin $u=v=0$ by
changing the value of the fluctuation parameters $u$ and $v$
(we must be on the second sheet to do so). 
Eqs.~(\ref{z2crit}) and (\ref{x2crit}) indicate a suppression 
of $X(2)$ and $Z(2)$, leading finally to their vanishing:
\begin{equation} \label{vanish2}
Z(2)\geq 0 \leftrightarrow u\geq 4kg\,, \quad
X(2)\geq 0 \leftrightarrow u\geq 4\sqrt{kf}\,,
\end{equation}
Which order parameter vanishes first depends on the relative size of $f$ and $g$:
if $g\geq \sqrt{f/k}$, $Z(2)$ vanishes before $X(2)$ does (and the other way round otherwise).
The point where at least one of the conditions Eq.~(\ref{vanish2}) is fulfilled marks the endpoint
of the critical line : if we could proceed further down the critical line,
we would end up with an unphysical situation where $X(2)$ or $Z(2)$ is negative. 

One can check that the two conditions in Eq.~(\ref{vanish2})
are equivalent to the paramagnetic inequalities Eq.~(\ref{paramuv}) along the critical 
line $z=0$ -- where $y=4/u=2/\sqrt{v}$ and $SU(n+2)\times SU(n+2)$ order parameters
vanish. This was expected, since the paramagnetic inequalities along
the critical line yield $X(2)\geq X(n+2)=0$ and $Z(2) \geq Z(n+2)=0$, i.e.
reduce to positivity constraints for the two-flavour order parameters.
On the second sheet, the critical line ends thus at 
the edge of the domain allowed by the paramagnetic inequalities. 
In the physical case $n=1$, and for $r=25$, the right hand side of Fig.~\ref{fig:paramag}
shows that the inequality for $v$ (i.e. for $X$) is saturated first
and that the endpoint of the critical line corresponds to $X(2)=0$, $Z(2)>0$.
From App.~\ref{app:fg}, we see that this occurs in the physical case for any (large) 
value of $r$.

In this section, we have investigated the critical line of $SU(n+2)\times SU(n+2)$ 
symmetry restoration. Along this critical line, 
both $Z(n+2)$ and $X(n+2)$ vanish, while $Y(n+2)=X(n+2)/Z(n+2)$
remains non-zero and finite. 
We have studied the two-flavour order parameters $X(2)$ and $Z(2)$ as well. 
In the regions
of the two sheets allowed by the paramagnetic inequalities, both are different from 0.
There is only one exceptional point on the second sheet, where the critical line and the most
stringent paramagnetic bound intersect. At this endpoint of the critical line, one of the two
$SU(2)\times SU(2)$ order parameters vanishes -- $X(2)$ in the physical case $n=1$.

\section{No-fluctuation limit: $N_c \to \infty$}
\label{sec:7}

We briefly sum up in this section the properties of another interesting region in the
$(u,v)$ plane where the effect of vacuum fluctuations is suppressed. This
can be realised as the large-$N_c$ limit of the theory. Since
\begin{equation} \label{uvnc}
u = O(1/N_c) \,, \quad v = O(1/N_c) \,,
\end{equation}
we deal with the vicinity of the origin on the first sheet. The large-$N_c$
limit forces the perturbative solution of the generic system Eqs.~(\ref{u})-(\ref{v}):
\begin{equation} \label{yznc}
y = 1 + O(1/N_c) \,, \qquad z = 1 + O(1/N_c) \,.
\end{equation}
The analogy between $1/N_c$ expansion and $\chi$PT ceases here. The large-$N_c$ limit does
not have much to say about the expansion of the rescaling factors in Eqs.~(\ref{yzrescaled}) and
(\ref{uvrescaled}) -- except perhaps by providing estimates of higher-order counterterms
included in the NNLO remainders $d_n$ and $e_n$.
Combining the result (\ref{yznc}) with Eqs.~(\ref{yzrescaled}) and (\ref{uvrescaled}) one
gets for large $N_c$:
\begin{equation} \label{ordernc}
X(n+2)\to 1-\epsilon(r)-d_n\,, \qquad
Z(n+2)\to 1-\eta(r)-e_n\,,
\end{equation}
From Eqs.~(\ref{xtwo}) and (\ref{ztwo}), we are now able to infer the 
two-flavour order parameters $X(2)$ and $Z(2)$ in the large-$N_c$ limit. One of
course expects to reproduce Eq.~(\ref{ordernc}) --
any dependence on the number of flavours should disappear
at the leading order in $1/N_c$. One first remarks that:
\begin{equation}
f_n = O(1/N_c) \,, \qquad g_n = O(1/N_c) \,,
\end{equation}
so that for $N_c\to\infty$ the right hand sides of Eqs.~(\ref{ztwo}) and
(\ref{ztwo}) both coincide with the one given in Eq.~(\ref{ordernc}). It remains to
show that the NNLO remainders $\bar d_{\pi n}=\lim_{m \to 0}d_{\pi n}$ and
$\bar e_{\pi n} =\lim_{m \to 0}e_{\pi n}$ are also suppressed
for large $N_c$. Indeed, the large-$N_c$ counting of connected QCD correlation
functions yields:
\begin{equation}
\lim_{N_c \to \infty} \frac{\partial}{\partial m_s} \log (F_\pi^2 M_\pi^2)= 0 \,,\quad
\lim_{N_c \to \infty} \frac{\partial}{\partial m_s} \log (F_\pi^2)= 0 \,,
\end{equation}
order by order in powers of quark masses, since the involved quantities
receive contributions only from connected graphs with two (and more) fermion loops. 
This shows the suppression of  $\bar d_{\pi n}$ and $\bar e_{\pi n}$
and thus the desired result:
\begin{equation}
X(n+2) = X(2) + O(1/N_c) \,, \quad Z(n+2) = Z(2) + O(1/N_c) \,.
\end{equation}
Both the condensate and the decay constant are independent of the number of
massless flavours, and for not too small $r$ both $X$ and $Z$ are close to
$1$. For completeness, we mention the leading large-$N_c$ behaviour of the
LEC's $L_5(\mu)$ and $L_8(\mu)$, which can be read off from Eqs.~(\ref{L8}) and
(\ref{L5}):
\begin{eqnarray}
L_5 &=&\frac{F_\pi^2}{8 M_\pi^2 } [\eta(r)+ e'_n] [1-\eta(r) - e_n] k_n(r)\,, \\
L_8 &=& \frac{F_\pi^2}{16 M_\pi^2} [\epsilon(r) + d'_n] [1- \epsilon(r) - d_n]k_n(r)\,.
\end{eqnarray}
Notice that the scale dependence only shows up at the next-to-leading
order. The OZI-rule suppressed constants $L_4$ and $L_6$ remain $O(1)$
at large $N_c$. They determine how quickly the fluctuation parameters
vanish as $N_c$ increases.

\section{The limit of large fluctuations} \label{sec:largefluc}
\label{sec:8}

                            We have seen that both multi-flavour
($N_f\ge 3$) $\chi$PT and $1/N_c$ expansion require small fluctuation parameters
$u$ and $v$. On the other hand, in the real world with three light flavours
($n = 1$), the fluctuation parameters $u_1$, $v_1$, as well as the
difference $u_1 - v_1$, are not small compared to 1. Sum-rule 
studies~\cite{Bachir1,Bachir2,D} suggest a
significant though small deviation of the LEC's $L_6(\mu)$ and $L_4(\mu)$
from their critical values Eqs.~(\ref{ncritical4}) and (\ref{ncritical6}). The large
coefficients in Eq.~(\ref{lambdarho}) then amplify this deviation resulting into
fluctuation parameters well above 1.  We will now study the effect
of large fluctuations on the pattern of chiral symmetry breaking.

\subsection{Many-flavour limit}
\label{sec:8.1}

QCD does not contain an obvious parameter which could allow one to
describe the limit of large fluctuations. Indeed, in the limit $n\to
\infty$, the effect of sea-quark pairs, such as the induced condensate
or the OZI-rule violation in the vacuum  channel, are
enhanced. However, a large-$N_f$ limit with
$N_c$ held fixed would presumably meet phase transitions,
leading eventually to the restoration of chiral symmetry and the loss
of asymptotic freedom. The definition of a large-fluctuation limit at
the level of QCD is thus likely to require a combined limit of
large $N_f$ and large $N_c$. However, at the level of the identities
of the effective theory, we are allowed to take the formal limit $n\to
\infty$. We will see  that this limit provides a natural example of
large fluctuations, if we make the assumption that in this limit the
pseudoscalar masses and decay constants remain finite order by order
in powers of quark masses $m$ and $m_s$.

Indeed, within such assumption, the fluctuation
parameters $\rho$ and $\lambda$ grow with $n$, unless the constants
$L_4^n(\mu)$  and $L_6^n(\mu)$ are set to the asymptotic values of
$L_4^{n,\mathrm{crit}}$ and $L_6^{n,\mathrm{crit}}$:
\begin{equation}\label{as}
L_4^n(\mu) \to L_4^{\mathrm{as}}(\mu), \quad L_6^n(\mu) \to
L_6^{\mathrm{as}}(\mu),
\end{equation}
where
\begin{equation}\label{46as}
L_4^{\mathrm{as}}(\mu) = 2 L_6^{\mathrm{as}} (\mu) = \frac{1}{256
\pi^2} \left( \log \frac{M_K^2}{\mu^2} - \frac{2}{r-1} \log
\frac{M_X^2}{M_K^2} \right) .
\end{equation}
From the chiral expansion of the masses and decay constants,
Eqs.~(\ref{pinmass})-(\ref{kanmass}) and
(\ref{pindecay})-(\ref{kandecay}), and assuming that they remain
finite order by order in powers of $m$ and $m_s$, one can read   the
a-priori unknown large-$n$ 
behaviour of  $Y(n+2)$, $Z^s_n$, $\xi_n$, $\tilde \xi$ and $A_n$,
\begin{equation}
\label{YZxit}
Y(n+2) \sim  Z^s_n \sim \tilde \xi_n \sim 1/n, \quad A_n \sim  \xi_n
\sim 1.
\end{equation}
The constraint on $Y(n+2)$ is imposed by the presence of the logarithmic
term (absent in the physical case $n=1$):
\begin{equation} \label{unstab}
L'_n \sim -\frac{n}{16\pi^2} \log\frac{M_X^2}{M_K^2} \to -\infty\,.
\end{equation}
This implies in turn that $L^n_6$ should grow like $n$, whilst
 $L_4^n$ should approach a finite value, which need not coincide with
 the  critical asymptote~(\ref{46as}).  

Actually, the same large-$n$ behaviour of the theory is also imposed by
the paramagnetic inequalities, provided that $f_n$ and $g_n$, defined
in Eqs.~(\ref{deffn})  and (\ref{defgn}) are positive:
\begin{eqnarray}
f_n \sim g_n &\sim& \frac{nr}{32\pi^2}\frac{M_\pi^2}{F_\pi^2}
\left[\frac{2}{r-1}\log\frac{M_X^2}{M_K^2}+\log\frac{\bar{M}_K^2}{M_K^2}\right]
\nonumber \\
&=& n l_\infty \geq 0\,.
\end{eqnarray}
The paramagnetic inequality for $Z$ -- second inequality
in Eq.~(\ref{paramy}) -- implies that $y$ vanishes like $1/n$ (or more
quickly). The limit $y\to 0$ is allowed  only on the first sheet
($y_+$) and Eq.~(\ref{ysol}) leads to $v_n-u_n\to\infty$ like $n^2$
(or more quickly). 
On the other hand, the paramagnetic constraint for $u$ 
-- second inequality in Eq.~(\ref{paramuv}) -- leads to $u=O(n)$, unless
a cancellation between $kg$ and $E/y$ occurs. Such a cancellation between
the NNLO remainder $E$ and the one-loop chiral logarithms $g$ cannot be
logically ruled out. It would however stand against our initial assumption 
that the chiral expansions exhibit a good overall convergence starting
at the NNLO order independently of $n$.

Let us now describe more precisely the large-$n$ behaviour of
$u$ and $v$. Since $u$ is positive, $z=1-uy/4$ cannot diverge.
Let us call $z_\infty=\lim_{n\to\infty} z_n$ and introduce
 a second parameter $a$ such that:
\begin{equation} \label{fluclimit}
u\sim\frac{4n}{a}(1-z_\infty)\,, \quad v\sim 4n^2/a^2\,,
\quad  y\sim a/n\,.
\end{equation}
$a$ and $z_\infty$ describe the behaviour of the fluctuation parameters $u$ 
and $v$. 
The order parameters tend to finite values:
\begin{equation}
\begin{array}{rcl}
X(n+2)&\to& 0\,, \\
Z(n+2)&\to& (1-\eta(r)-e_n) z_\infty\,,\\
X(2)[1-\bar{d}_{\pi n}] &\to& 1-\epsilon(r)-d_n\,,\\
Z(2)[1-\bar{e}_{\pi n}] &\to&  [1-\eta(r)-e_n][1-akl_\infty]\,.
\end{array}
\label{largenorder}
\end{equation}
In this limit, the LEC's associated with the decay constant identities 
remain scale-dependent -- $L_5=O(n)$ and $L_4=O(1)$ -- whereas the LEC's arising
in the mass identities become:
\begin{eqnarray}
L_6^n(\mu) &\sim & \frac{n}{16rka^2}\frac{F_\pi^2}{M_\pi^2}\,,\label{largenl6}\\
L_8^n(\mu) &\sim & \frac{n^2}{16 a^2}
   \left(\frac{1-\eta(r)-e_n}{1-\epsilon(r)-d_n}\right)^2(\epsilon(r)+d'_n)
     \frac{F_\pi^2}{M_\pi^2}\,.
\label{largenl8}
\end{eqnarray}
A comment is in order here about the double limit of large
$N_c$ and $N_f$, investigated in Ref.~\cite{GST}. 
If we consider the large-$n$ formulae for the LEC's $L_{4,5,6,8}^n$
[see Eqs.~(\ref{largenl6})-(\ref{largenl8})], we can recover their standard
behaviour in the large-$N_c$ limit,
provided that $N_c\sim n$ and $a=O(N_c)$, i.e. $a=F_\pi^2/\Lambda^2$ with
$\Lambda$ an $N_c$-independent scale~\footnote{
Needless to say that the
``holomorphic phase'' analysed in Ref.~\cite{GST} is outside the scope of the
present discussion which is exclusively based on an effective theory
describing the breakdown $ SU(N_f) \times SU(N_f) \to SU_V(N_f)$.}.

The large-$n$ limit of the theory illustrates therefore
how the large-fluctuation limit can be formulated consistently:
the multi-flavour quark condensate is then
suppressed whereas the two-flavour condensate $X(2)$ remains
different from zero (and close to 1 for $r\geq 15$).

\subsection{Large fluctuation parameters in three-flavour QCD}

The many-flavour limit of the theory discussed above
should merely be viewed as a particular realisation of the limit of large fluctuations which is hopefully consistent but not necessarily unique. The limit of large fluctuations could as well be formulated directly in terms
of fluctuation and order parameters keeping the number of flavours fixed to
$N_f=3$. In that case ($n = 1$), one avoids the presence of extra $n^2 - 1$ Goldstone
boson X-states arising for $n > 1$. Since the limit is designed in a slightly
different way from the previous section, we are ending up with similar but not identical
results for the LEC's and the two-flavour parameters.

One may infer from the generic Eqs.~(\ref{u}) and (\ref{v}) the lines
in the $(u,v)$ plane corresponding to a constant value of $z=z_1$ ($ 0<z_1<1 $):
\begin{equation}
u = \frac{4}{y}(1 - z_1) \,, \quad v = \frac{4}{y^2} (1 - yz_1)\,.
\end{equation}
It is seen that keeping $z$ fixed and setting $y \to 0$ one reproduces the
large-fluctuation behaviour Eq.~(\ref{fluclimit}). The decay constant
$ Z(3) = [1 - \eta(r) - e_1] z_1 $ remains nonzero, whereas the three-flavour 
condensate vanishes asymptotically:
\begin{equation}
X(3) = [1 - \eta(r) - e_1] z_1 y  \to 0 \,.
\end{equation}
In this large-fluctuation limit for $n=1$, all relevant LEC's appearing
in the original mass and decay constant identities can be predicted in terms
of $ r = m_s/m $ and of the parameter $ Y(3) = X(3) / Z(3)  \to 0 $. The
leading behaviour of $L_5$ and $L_8$ is then:
\begin{eqnarray}
L_5&\sim& \frac{F_\pi^2}{8M_\pi^2}\frac{\eta(r)+e'_1}{Y(3)},
\,\\
L_8&\sim& \frac{F^2_\pi}{16M^2_\pi} \frac{\epsilon(r) + d'_1}{Y^2(3)}\,.
\end{eqnarray}
Notice that the OZI-rule violating constant $L_6(\mu)$ is no more suppressed
[to be compared with the large-$n$ result Eq.~(\ref{largenl6})]. 
This is best seen from the ratio $L_6/L_8 $ which in the
$y \to 0$ limit becomes:
\begin{equation}\label{ratio6}
\lim_{y \to 0} \frac{L_6(\mu)}{L_8(\mu)} = \frac{1 - \epsilon(r) - d_1}
{(r + 2)[\epsilon(r) + d'_1]}\,.
\end{equation}
It amounts to: $L_6/L_8 = 0.43,\, 0.79,\, 1.53$ for $ r = 20,\, 25,\, 30$
respectively. The second OZI-rule violating ratio $L_4/L_5$ still depends
on the order parameter $Z(3)$:
\begin{equation}\label{ratio4}
\lim_{y \to 0} \frac{L_4(\mu)}{L_5(\mu)} = \frac{1 - \eta(r) - Z(3) - e_1}
{(r + 2)[\eta(r) + e'_1]}\,.
\end{equation}
This ratio could be more suppressed provided $Z(3)$ is close to 1.
In the latter case the expression ~(\ref{ratio4}) could be rather sensitive
to the NNLO remainder $e_1$, contrary to the case of the ratio Eq.~(\ref{ratio6})
where $d_1$ competes with 1 (recall that both $d'_1$ and $e'_1$ are
suppressed by an extra power of $1/r$).

                      It remains to work out the two-flavour order parameters
$X(2)$ and $Z(2)$ in the $y \to 0$, $n=1$ large-fluctuation limit:
\begin{eqnarray}
X(2)[1-\bar{d}_{\pi 1}] &=& (1-\epsilon(r) - d_1) \frac{r}{r+2}\,,\\
Z(2)[1-\bar{e}_{\pi 1}] &=& (1-\eta(r)-e_1) \frac{r+2z_1}{r+2}\,.
\end{eqnarray}
Despite the vanishing of the three-flavour condensate $X(3)$ $\to 0$,
the two-flavour condensate $X(2)$ remains non zero and close to 1,
provided $r=m_s/m$ is not too small. This effect is entirely due to
the induced condensate and it is proportional to the strange quark mass.
It is worth stressing the fundamental difference between the chiral
symmetry restoration which occurs along the critical line $v = u^2/4$
for finite $u$ and $v$ and the large-fluctuation limit in which
$ (u,v) \to \infty$. This difference merely occurs for three-flavour order
parameters:
whereas in the symmetry restoration case both $Z(3)$ and $X(3)$ vanish
and their ratio $Y(3)=X(3)/Z(3)$ remains non-zero, the large fluctuation
limit is characterised by a continuous decrease of $Y(3)$ and of the
condensate, with the decay constant $Z(3)$ held fixed. This manifestation
of large fluctuations need not correspond to a phase
transition: they would lead naturally to a spontaneous breakdown of
chiral symmetry with a very small (but non-vanishing) quark condensate.
In this context, the recently proposed ``no-go'' theorems~\cite{Shifman} 
-- stating that the vanishing of the condensate 
implies a vanishing pion decay constant and chiral symmetry 
restoration -- do not necessarily apply.

\begin{table*}[ht]
\caption{Values of chiral order parameters in three different limits ($n=1$). The
second column
corresponds to small fluctuations (large-$N_c$ limit). The third
one stands for the large-fluctuation limits ($y\to 0$, $z$ set to $z_1$).
The fourth one indicates chiral symmetry restoration ($z\to 0$, $y$ set to
$y_1$).\label{tab:yfluc}}
\begin{center}
\begin{tabular}{|c|c|c|c|}
\hline
Limit & $N_c\to\infty$ & Large fluct. $(y\to 0)$ & $\chi$ rest. $(z\to 0)$\\
\hline
$u_1$ & 0 & $4(1-z_1)/y$    & $4/y_1$\\
$v_1$ & 0 & $4(1-yz_1)/y^2$ & $4/y_1^2$\\
\hline
$Y(3)$ & $\displaystyle \frac{1-\epsilon(r)-d_1}{1-\eta(r)-e_1}$ 
       & 0
       & $\displaystyle\frac{1-\epsilon(r)-d_1}{1-\eta(r)-e_1} y_1$\\
\hline
$Z(3)$ & $1-\eta(r)-e_1$ 
       & $z_1[1-\eta(r)-e_1]$
       & 0\\
&&&\\
$Z(2)$ & $1-\eta(r)-e_1$  
       & $\displaystyle\frac{r+2z_1}{r+2}\frac{1-\eta(r)-e_1}{1-\bar{e}_{\pi 1}}$
       & $\displaystyle\frac{r}{r+2}
          \frac{1-\eta(r)-e_1}{1-\bar{e}_{\pi 1}}[1-y_1k_1(r)g_1]$ \\
\hline 
$X(3)$ & $1-\epsilon(r)-d_1$
       & 0
       & 0\\
&&&\\
$X(2)$ & $1-\epsilon(r)-d_1$ 
       & $\displaystyle\frac{r}{r+2}\frac{1-\epsilon(r)-d_1}{1-\bar{d}_{\pi 1}}$
       & $\displaystyle\frac{r}{r+2}\frac{1-\epsilon(r)-d_1}{1-\bar{d}_{\pi 1}}
           [1-y_1^2 k_1(r)f_1]$ \\
\hline
\end{tabular}
\end{center}
\end{table*}
\begin{figure}[t]
\centerline{
\epsfysize8cm\epsffile{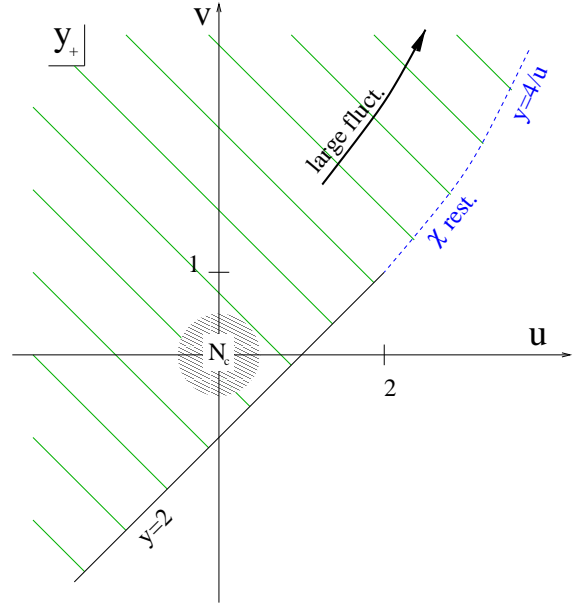}
}
\caption{Limits of small and large fluctuations, and of chiral symmetry restoration
on the $y_+$ sheet for $n=1$:
the shaded region around the origin denoted $N_c$ corresponds to the
domain of application for the large-$N_c$ limit, the arrow illustrates
the large-fluctuation limit $u,v\to\infty$, and the ``$\chi$ rest.'' border
is the critical line of chiral symmetry restoration where $z=0$.
\label{fig:yfluc}}
\end{figure}

The limits of small and large fluctuations and of chiral symmetry restoration
are summarised in  Fig.~\ref{fig:yfluc}, while the corresponding results 
for the order parameters are collected in Table~\ref{tab:yfluc}.

\section{Two-flavour Chiral Perturbation Theory} \label{sec:twoflav}  
\label{sec:9}

                     The striking outcome of the previous analysis is the
persistence of a large two-flavour condensate $\Sigma(2)$ even
when the multi-flavour condensate $\Sigma(N_f)$ $[N_f \ge 3]$ is suppressed.
We have seen that $X(3)$ could be well below 1, indicating that the
expansion of $F^2_\pi M^2_\pi$ in powers of $m_u, m_d$ and $m_s$ need not be
dominated by the genuine condensate term $(m_u+m_d) \Sigma(3)$. Actually, the
whole $N_f \ge 3$  $\chi$PT treated in the standard way can exhibit
instabilities, not because of too large a strange-quark mass, but rather
because of too sizeable vacuum fluctuations of $\bar{q}q$ pairs.
On the other hand, the expansion of the same quantity
$F^2_\pi M^2_\pi$  in powers of $m_u , m_d$ only is expected to be dominated
by the $N_f = 2$ condensate term  $(m_u+m_d) \Sigma(2)$.
Indeed, for not too small $r = m_s/m > 15$, the result  $X(2) \sim 1$
is emerging independently of the fluctuation parameters $u$ and $v$
reflecting an important contribution of the induced condensate.
This suggests that for suitable pion observables, the standard
$SU(2) \times SU(2)$ $\chi$PT~\cite{GL1}
could be a well-convergent expansion scheme.
                       
		       In order to gain more insight into the different
behaviour of two- and multi-flavour chiral dynamics, it is convenient to
rewrite the $N_f=2$ Ward identities generating the expansion of
$F^2_\pi M^2_\pi$ and of $F^2_\pi$ in a form as close as possible to
Secs.~\ref{sec:model} and \ref{sec:connect}. 
They involve the condensate $\Sigma(2)$, the
decay constant $F^2(2)$ and the two $O(p^4)$ symmetry-breaking
scale-independent LEC's $\bar{\ell}_3$ and $\bar{\ell}_4$~\cite{GL1}:
\begin{eqnarray}\label{mass2}
F^2_\pi M^2_\pi &=& 2m\Sigma(2) + 
   \frac{m^2B^2(2)}{8\pi^2}(4 \bar{\ell}_4 - \bar{\ell}_3) +
F^2_\pi M^2_\pi \delta\,,\\
\label{decaycst2}
F^2_\pi &=& F^2(2) + \frac{mB(2)}{4\pi^2}\bar{\ell}_4 + F^2_\pi \varepsilon\,.
\end{eqnarray}
Here $B(2)= \Sigma(2)/F^2(2)$ and
the NNLO remainders $\delta$ and $\varepsilon$~\footnote{The remainder
$\varepsilon$ should not be confused with the function $\epsilon(r)$     
introduced in Section 3.2.} are $O(m^2)$, expected to be of order 1\%. 
The analogy with the multi-flavour case can be pushed
further by rewriting Eqs.~(\ref{mass2}) and (\ref{decaycst2}) in the form
of Eqs.~(\ref{xl6})-(\ref{zl4}):
\begin{equation}
X(2) =1-\delta-Y(2)^2 \bar\rho/4 \,, \qquad Z(2)=1-\varepsilon-Y(2)\bar\lambda/4\,,
\end{equation}
where:
\begin{equation}
\bar\rho = \frac{1}{8\pi^2} \frac{M^2_\pi}{F^2_\pi} (4\bar\ell_4-\bar\ell_3)\,, \quad
\bar\lambda = \frac{1}{8\pi^2} \frac{M^2_\pi}{F^2_\pi} 4 \bar\ell_4 \,.
\end{equation}
The simple rescaling:
\begin{equation}
\bar{z}=\frac{1}{1-\varepsilon} \frac{F^2(2)}{F^2_\pi} \,, \quad
\bar{y}=\frac{1-\varepsilon}{1-\delta} \frac{2mB(2)}{M^2_\pi} \,, 
\end{equation}
\begin{equation}
\bar{u} = \bar\lambda \bar{k} \,, \quad \bar{v} = \bar{\rho} \bar{k} \,, \quad 
\bar{k} = \frac {1- \delta}{(1- \varepsilon)^2}\,,
\end{equation}
brings the two-flavour mass and decay constant identities into the standard
form (\ref{u})-(\ref{v}):
\begin{eqnarray}\label{fund2}
\bar{z} + \frac{1}{4} \bar{u}\bar{y} &=& 1 \,, \\
\bar{z} + \frac{1}{4} \bar{v}\bar{y} &=& \frac{1}{\bar{y}}\,. \label{fundz2}
\end{eqnarray}
The formal analogy is now complete. In the two-flavour case the factors
$\epsilon(r)$ and $\eta(r)$ have to be omitted in the rescaling factors
but otherwise, all equations look identical. Where is the difference?

The multi- and two-flavour cases behave differently
because the corresponding parameters $v$ and $u$ (or $\rho$ and $\lambda$)
are respectively of a different origin and magnitude. We know from the
previous discussion that the two-flavour quantity $\bar{y} \sim 1$ and
consequently, the parameters $\bar{v}-\bar{u}$ and $\bar{u}$ cannot be too large.
This in turn excludes an unlimited grow of $|\bar\ell_3|$. As a consequence
one may expect the perturbative solution of the fundamental
equations~(\ref{fund2}) and (\ref{fundz2}) [obtained 
by Taylor expanding the non-perturbative solution] to
make sense.

We have already emphasised the connection of the multi-flavour parameters
$v_n , u_n$ $(n \geq 1)$ with the correlations between
vacuum strange and non-strange $\bar qq$ pairs and (last but not least) with
the fluctuations of small Euclidean Dirac eigenvalues~\cite{DGS}. We do not
know much about the importance of these fluctuations from first principles,
but we do understand why in the $N_f \geq 3$ theory such fluctuations
manifest themselves through important OZI-rule violations in the vacuum channel
$J^P = 0^{+}$. Hence, if the quark pairs in the vacuum are strongly
correlated and/or the low-energy Dirac spectrum is subjected to large
fluctuations, the multi-flavour parameters $ u_n , v_n$ are likely large
and the perturbative solution of the corresponding fundamental
equations~(\ref{u}) and ~(\ref{v}) breaks down.
On the other hand, in a $N_f = 2$ theory and in the presence of massive
$(m_s \sim \Lambda_{QCD})$ strange quarks in the sea,
the same cause does not produce the same effect.
In this case, the fluctuations of
small Dirac eigenvalues are much harder to relate to low-energy
observables: the OZI-rule is inoperative in this case, and the scalar
correlator $\langle (\bar uu) (\bar dd)\rangle$ is chirally invariant and
not simply related to an observable order parameter. The different nature
of parameters $\bar{u}$ and $\bar{v}$ is further illustrated
by a different behaviour in the large-$N_c$ limit:
whereas the multi-flavour fluctuation parameters
$u_n$, $v_n$ for $n \ge 1$ are suppressed as $O(1/N_c)$, $\bar{u}$ and $\bar{v}$
behave as $O(1)$ since the constants $\bar\ell_3$ and $\bar\ell_4 $ 
behave like $O(N_c)$.

The fact that $\bar{v}, \bar{u}$ (or $\bar\rho,\bar\lambda$) as  well as the LEC's 
$\bar\ell_3, \bar\ell_4$ need not be enhanced by fluctuations of small
Dirac eigenvalues can be seen directly by comparing (matching) the multiflavour
mass identity (\ref{pinmass}) with its $N_f=2$ counterpart (\ref{mass2}).
In the former, the vacuum-fluctuation contribution resides in the term
$Z_n^s$ containing the LEC $L_6(\mu)$. Eq.~(\ref{mass2}) is a 
reexpression of the same mass identity in terms of        
$SU(2)\times SU(2)$ quantities. Both identities have to match
order by order in $m$. At the first order, one obtains the expression for
the two-flavour condensate:
\begin{equation}
\Sigma(2) = \Sigma(n+2) +n m_s \bar Z_n^s + \Sigma(2)\bar{d}_{\pi n}\,,
\end{equation}
which absorbs the major part (the term proportional to $m_s$)
of the vacuum fluctuation contribution $(nm_s+2m) Z^s$. 
$n m_s \bar Z_n^s$ is what we have called the induced 
condensate~\cite{DGS,GST}. 
The remaining part of $Z^s$ contributes together with
$A_n$ (i.e. $L_8$) to the $N_f=2$ LEC's $\bar\ell_3$ and $\bar\ell_4$, but 
this contribution contains neither $m_s$ nor the flavour factor $n$.
Consequently, there is no particular reason why fluctuations of small Dirac
eigenvalues should enhance the parameters $\bar{v}$ and $\bar{u}$ and to destabilise
the two-flavour standard $\chi$PT.             

                        The non-perturbative effects in the non-linear system
(\ref{fund2})-(\ref{fundz2})
start to show up for $|\bar{v}-\bar{u}| \sim 1 $ corresponding to
$|\bar\ell_3| \sim 35 $. At this point, it is worth stressing that the
original standard estimates of $\bar\ell_3$, $\bar\ell_4$ ~\cite{GL1,GL2}  
make use
of $N_f=3$ observables assuming the approximate validity of
the OZI-rule in the scalar channel. Since vacuum fluctuations do
contribute to $\bar\ell_3,\, \bar\ell_4$, the standard estimates~\cite{GL1,GL2} 
$\bar\ell_3 =2.9\pm 2.4$ and  $\bar\ell_4 = 4.3 \pm 0.9$     
could be modified by OZI-rule violating effects.

                       In order to relate $\bar\ell_3,\, \bar\ell_4$ to
$N_f=3$ observables including OZI-rule violating vacuum fluctuations, one may
proceed as follows. First the two-flavour mass and decay constant identities
Eqs.~(\ref{mass2}) and (\ref{decaycst2}) are equivalently rewritten to express
$\bar\ell_3 , \bar\ell_4$ in terms of the order parameters $X(2)$ and $Z(2)$ :
\begin{eqnarray}
\bar\ell_3 &=& 32 \pi^2 \frac{F^2_\pi Z(2)}{M^2_\pi X(2)}\left[1-\varepsilon
- (1-\delta)\frac{Z(2)}{X(2)}\right ] \,, \\
\bar\ell_4 &=& 8 \pi^2 \frac{F^2_\pi Z(2)}{M^2_\pi X(2)}[1 - \varepsilon -
Z(2)]\,.
\end{eqnarray}
Next, one uses Eqs.~(\ref{x2simple}) and (\ref{z2simple})      
relating X(2) and Z(2) to the multi-flavour order
parameters $y_n$ and $z_n$. In the case of three flavours ($n=1$) the
latter may be written as:
\begin{eqnarray} \label{z2match}
Z(2)
 &=&\frac{r}{r+2} [1-\eta(r)-(e_1 - \bar{e}_{\pi 1}\ldots )] \nonumber \\
 & & \times
    \left\{1-y_1 k_1(r) g_1 + \frac{2}{r} z_1\right\}\,,\\
X(2)
 &=&\frac{r}{r+2} [1-\epsilon(r)-(d_1 - \bar{d}_{\pi 1}\ldots )] \nonumber \\
 & & \times \left\{1-y_1^2 k_1(r) f_1 + \frac{2}{r} y_1 z_1\right\}\,, \label{x2match}
\end{eqnarray}
where the ellipses stand for terms of order $(\bar e_{\pi 1})^2$,
$(\bar d_{\pi 1})^2$ as well as $\bar e_{\pi 1} \eta(r)$ and
$\bar d_{\pi 1}\epsilon(r)$. Since $\bar e_{\pi 1},\bar d_{\pi 1}=O(m_s^2)$
are expected of order $0.1$ or less, the dotted terms in Eqs.~(\ref{z2match})
and (\ref{x2match}) can hardly exceed 1 \%. Furthermore, the leading $O(m_s^2)$
terms cancel out in the differences $e_1-\bar e_{\pi 1}$ and
$d_1-\bar d_{\pi 1}$. As a result, the whole contribution of NNLO remainders
in Eqs.~(\ref{z2match}) and (\ref{x2match}) is $O(m m_s)$, i.e. they can be
expected of order 3 \%.

                  It is instructive to exploit the above relations and to
estimate the order parameters $X(2), Z(2)$ and 
the LEC's $\bar\ell_3, \bar\ell_4$ that correspond 
to the extreme cases of no fluctuation
($N_c \to \infty$) and large fluctuations ($y_1 \to 0$). Neglecting all
NNLO remainders, one gets for $N_c \to \infty$:
\begin{equation}
r=20: \quad \begin{array}{cc} X(2)=0.905\,,&  Z(2)=0.949\,, \\ \bar\ell_3=-7.5\,,
& \bar\ell_4=2.0\,, \end{array} \end{equation}
\begin{equation}
r=25:  \quad \begin{array}{cc} X(2)=0.955\,,& Z(2)=0.959\,, \\ \bar\ell_3=-0.6\,, 
& \bar\ell_4=1.5\,, \end{array}
\end{equation}
These estimates are quite compatible with standard 
expectations~\cite{GL1,GL2}. On the other hand, one gets in the limit of
large fluctuations (taking $z_1 = 0.7$):
\begin{equation}
r=20: \quad \begin{array}{cc} X(2)=0.822\,, & Z(2)=0.922\,, \\
\bar\ell_3=-20.2\,,& \bar\ell_4=3.2\,,\end{array}
\end{equation}
\begin{equation}
r=25: \quad \begin{array}{cc} X(2)=0.884\,,& Z(2)=0.938\,, \\
\bar\ell_3=-9.6\,, &  \bar\ell_4=2.4\,,\end{array}
\end{equation}  

As expected, large vacuum fluctuations push $\bar\ell_3$ towards larger
negative values, without however reaching the range in which the two-flavour
$\chi$PT would start to diverge. The above estimates should be taken with caution:
even the small remainders in Eqs.~(\ref{z2match}) and (\ref{x2match})
of the order of a few per cent can give an important contribution to
$\bar\ell_3$ and $\bar\ell_4$ since the latter involve $X(2) - Z(2)$ and $1 - Z(2)$
respectively.

           The  information on low-energy $\pi\pi$
phases extracted from  the new E865 Brook\-haven $K_{e4}$ experiment~\cite{E865} combined
with older data on $I=0$ and $I=2$ S-waves~\cite{pipidata}, 
have been recently used to extract
the two-flavour order parameters $X(2),Z(2)$ as well as the LEC's
$\bar\ell_3,\bar\ell_4$ in a model-independent analysis~\cite{DFGS} which is merely
based on the numerical solution of Roy equations~\cite{ACGL}. The result of 
this analysis reads:
\begin{eqnarray}
X(2) = 0.81 \pm 0.07\,, &\quad &  Z(2) = 0.89 \pm 0.02 \,, \\
\bar\ell_3 = -17.8 \pm 15.3\,, &\quad & \bar\ell_4 = 4.1 \pm 0.9 \,,
\end{eqnarray}
It is compatible with the estimates given above, and especially with 
the ones corresponding to the large-fluctuation limit.

We would like to close this section with a comment
emphasising the importance of the new precise $\pi\pi$-scattering data
recently published~\cite{E865}, forthcoming~\cite{diracexp} or expected in a
near future~\cite{NA48}. Some time ago it has been pointed out~\cite{GChPT}
that no experimental test of the actual size of the quark condensate and of
the magnitude of the quark mass ratio $r = m_s/m$ was available.
No particular mechanism of $\chi$SB was anticipated at that time
and a general (less predictive) expansion scheme (Generalized
Chiral Perturbation Theory or G$\chi$PT) was proposed to analyse data
in a model-independent way. Furthermore, there was no or little
experimental evidence for the substantial violation of the OZI 
rule in the scalar channel that supports today the idea of 
important vacuum fluctuations and a sizeable difference 
between the two- and multi-flavour condensates. 

On the other hand, the present conclusion that $X(2)$
is close to 1 and its consequences for the standard two-flavour $\chi$PT
might appear at first sight as a result that we could have anticipated
using nothing but theoretical arguments similar to those in this
paper -- contradicting therefore the claims made in Ref.~\cite{GChPT}.
As a matter of fact, this conclusion
can be drawn only if one gets extra (and independent) information
about the size of the quark mass ratio $r=m_s/m$, excluding
small values such as $r\sim 10$.
In the present paper, a larger value of $r$ was assumed from
the onset. If $r$ happened to be small, the factor
$\epsilon (r) = 2 (r_2 - r) / (r^2 - 1)$ would  get close to 1,
affecting the rescaling factors in 
Eqs.~(\ref{yzrescaled})-(\ref{uvrescaled}) and (\ref{Xyz}).
According to Eq.~(\ref{x2match}), $X(2)$ would then be suppressed
independently of the size of vacuum fluctuations. Because of the inequalities
$X(2)\ge X(3)\ge 0$, both $X(3)$ and $X(2)-X(3)$ would then be suppressed.
Hence, for small $r$, the vacuum fluctuations and the induced
condensate would enhance $X(2)$ less significantly,
implying a weaker violation of the OZI rule in the scalar channel.
In all cases, a strong correlation between $r$ and $X(2)$ [\emph{not} $X(3)$]
persists~\cite{DGS,DS}. Hence an
important result of the new and/or forthcoming $\pi\pi$ scattering experiments
is -- among other issues -- to 
put a lower bound on the quark mass ratio $r$, ruling out
small values of the two-flavour GOR ratio $X(2)$. A precise quantitative
statement of this lower bound is matter of a detailed $SU(3) \times SU(3)$
analysis of $\pi\pi$ scattering along the lines of the present paper.
Such an analysis is in progress~\cite{soon}.

\section{Conclusion}
\label{sec:10}

\indent

\emph{i)} Chiral order parameters may exhibit a strong dependence 
on the number of light flavours. Such a phenomenon can be interpreted
as an important effect of sea-quark pairs, which is related to a large 
violation of the OZI-rule observed in the scalar channel and reflects
significant fluctuations of the lowest modes of the Euclidean Dirac operator.
In this paper we have focused on the precise interplay between chiral 
order and fluctuations.

\emph{ii)}  We have highlighted the particular role of the 
strange quark, whose mass $m_s \sim \Lambda_{QCD}$ is light enough to populate
the vacuum with massive $\bar ss$-pairs, but heavy enough to have influence on
the properties of the $SU(2)\times SU(2)$ chiral limit
with $m_u,m_d\to 0$ and $m_s$ fixed at its physical value. 
In particular, the two-flavour condensate
$\Sigma(2) =- \lim_{m_u,m_d\to 0} \langle\bar u u\rangle$
does receive an extra contribution from the massive vacuum $\bar ss$-pairs
through the OZI-rule violating and $SU(2)\times SU(2)$
symmetry-breaking correlation $\langle(\bar uu)(\bar ss)\rangle$. 
The latter is referred to as the induced condensate.
Such contribution is absent in the $SU(3)\times SU(3)$
chiral limit $m_{u,d,s}\to 0$, since the remaining
quarks ($c$, $b$, $t$) are too heavy to contribute significantly to
vacuum fluctuations. This may lead to significant differences in the
manifestation of chiral symmetry breaking in the two- and three-flavour 
chiral limits.

\emph{iii)} To probe the effect of massive quark pairs on 
chiral symmetry breaking, 
we have introduced a theory with two ultralight
$u,d$ quarks and $n$ degenerate copies of the (massive) strange quark. QCD
can be recovered as the particular case $n=1$. Because of the
mass hierarchy $m_{u,d} \ll m_s \ll \Lambda_H$, two different chiral limits
can be considered in this model: the two-flavour limit, where only
$m_{u,d}\to 0$ but $m_s\neq 0$, and the multi-flavour limit, where all 
$N_f=n+2$ light masses vanish.

\emph{iv)} The multi-flavour case is characterised
by two fluctuation parameters which measure the violation of the 
OZI-rule in the scalar sector. They are related to the two large-$N_c$ 
suppressed $O(p^4)$ constants $L_6(\mu)$ and $L_4(\mu)$. We have shown that Ward 
identities yield two non-linear relations between the fluctuation 
parameters and the two basic multi-flavour order parameters: 
the quark condensate $X(N_f)=2m\Sigma(N_f)/(F_\pi^2M_\pi^2)$
and the Goldstone-boson decay constant $Z(N_f) = F^2(N_f)/F^2_\pi$ 
(where $F(N_f)$ is the pion coupling constant $F^2_\pi$ in the
chiral limit).
These relations are a direct consequence of Ward identities: 
all higher chiral orders (NNLO and
beyond) are absorbed into a finite multiplicative renormalisation (rescaling)
of the order and fluctuation parameters. The effect of this 
renormalisation remains small (i.e. rescaling factors $\sim 1$) provided 
that the chiral series globally converge from the NNLO order.

\emph{v)} Taking the $SU(2)\times SU(2)$ limit 
$m_u, m_d \to 0$ of
mass and decay constant Ward identities, one obtains the two-flavour 
condensate $X(2)$ and decay constant $Z(2)$ in the presence of $n$ copies 
of massive $s$-quark pairs in the sea. We can then compare  
multi- and two-flavour order parameters as functions of 
fluctuation parameters:
\begin{itemize}

\item Multi-flavour chiral symmetry is restored along a critical line
in the plane of fluctuation parameters. Along this line, 
the multi-flavour condensate $X(N_f)$ and decay constant $Z(N_f)$ both vanish,
but their ratio $X/Z$ stays finite and non-zero.
The two-flavour condensate
and pseudoscalar decay constant do not vanish in this limit, except in
one exceptional point corresponding to the end-point of the critical line.

\item In the case of small fluctuations, we recover 
a large-$N_c$, mean-field
type behaviour. The order parameters then do not depend on the number
of light flavours, and the quark condensate is the dominant signal of chiral
symmetry breaking.

\item In the opposite limit of large 
fluctuations, the multi-flavour quark condensate $X(N_f)$ tends to zero, but 
chiral symmetry remains spontaneously broken, since the decay constant
$Z(N_f)$ stays away from 0. 
Correspondingly, the two-flavour order parameters receive large contributions
from massive sea-quark loops, so that their size is naturally large, and close
to the one expected in the large-$N_c$ limit.
We would like to stress that the presence of large fluctuations is not
necessarily related to the occurrence of a phase transition: they could
constitute a feature of the dynamics of the theory, implying in turn a
naturally small size of the quark condensate. In this case the two-flavour
condensate would be exclusively made of the induced contribution from the
massive quark pairs in the vacuum.
\end{itemize}

\emph{vi)} The usual treatement of $N_f \ge 3$ $\chi$PT
considers fluctuation parameters as small, which requires a very precise 
fine-tuning of the $O(p^4)$ constants $L_6$ and $L_4$. 
On the other hand, the available information about the OZI-rule
violation in the scalar channel suggests that fluctuations can be
enhanced due to the large effect of sea quarks.
Such large fluctuations destabilise
chiral expansions by suppressing the lowest order of the series (three-flavour
quark condensate), but they need not spoil the overall convergence.
To cope with such a situation, we propose to treat the fluctuation parameters
(related to the OZI-rule violating LEC's $L_4$ and $L_6$) non-perturbatively
when expressing the parameters of the effective Lagrangian in terms of
observables. In practice it amounts to replacing the perturbative 
solution of the non-linear relation between order and fluctuation 
parameters (expressed as an expansion in powers of the latter) by the 
corresponding exact algebraic solution. In this way one can reexpress the 
low-energy constants $mB_0$, $F_0$, $L_5$, $L_7$ and $L_8$ in terms of the two 
fluctuation parameters, the quark mass ratio $r = m_s/m$ and higher $\chi$PT 
contributions arising through the (hopefully convergent) expansion of 
rescaling factors around 1. This procedure corresponds to a 
non-perturbative resummation of the fluctuations encoded in $L_4$ and $L_6$.

\emph{vii)} We have shown that vacuum 
fluctuations do not suppress the two-flavour condensate. The 
$N_f=2$ GOR ratio $X(2)$ remains close to 1 and the standard 
two-flavour $\chi$PT is likely a well-convergent expansion scheme, 
provided that the quark mass ratio $r$ is not too small. A lower bound 
on $r$ can be inferred from precise low-energy 
$\pi\pi$ scattering data.

\emph{viii)} 
Additional observables (Goldstone boson scattering, decay form factors\ldots)
would have to be analysed along similar lines, considering their
chiral expansion up to NNLO and making explicit the non-perturbative
dependence on the fluctuation parameters. We will illustrate more precisely
this point for $\pi\pi$ and $\pi K$ scattering parameters 
in a future publication~\cite{soon}.
Applying this method to a sufficiently large set of precisely measured
observables should allow one to pin down the size of vacuum fluctuations,
to disentangle the effect of massive $\bar{s}s$-pairs on the pattern
of two-flavour chiral symmetry breaking and to determine by how much 
three-flavour chiral order parameters are suppressed. This will eventually
lead to a better understanding of the spontaneous breakdown of chiral symmetry
and its dependence on the number and hierarchy of light quarks.

\begin{acknowledgement}

We would like to thank N.~H.~Fuchs for collaboration at an early
stage of this work. Various discussions with J.~Bijnens, U.-G.~Meissner,
B.~Moussallam and P.~Talavera have been highly appreciated. 
Work partially supported by EEC-TMR contract
ERBFMRXCT 98-0169 (EURODAPHNE). SD acknowledges support by
PPARC, through grant PPA/G/S/1998/00530.  LG acknowledges partial
support from European Program HPRN-CT-2000-00149.
\end{acknowledgement}

\appendix

\section{Goldstone boson masses in multi-flavour QCD} \label{app:etax}

We want to discuss in this section how to combine the Ward identities for
the masses and decay constants of the Goldstone bosons. We are going to see
that we need as inputs $M_\pi$, $M_K$, $F_\pi$, $F_K$ and $M_\eta$
to fix the values of the $O(p^4)$ LEC's as functions of $r$, $B_{0n}$ and $F^2(n+2)$,
and then to obtain the values of the remaining observables 
$F_\eta^2$, $M_X^2$ and $F_X^2$.

First, we have to spell out all the Ward identities concerning the masses 
and decay constants of the Goldstone bosons.
The masses and decay constant identities for the pion and kaon were given in 
Eqs.~(\ref{pinmass})-(\ref{kanmass}) and Eqs.~(\ref{pindecay})-(\ref{kandecay}).
The Ward identities for the extra $X$ states
can be expressed in the more practical form:
\begin{eqnarray} \label{fxmx}
\frac{F_X^2 M_X^2}{F_K^2 M_K^2}
  &=&\frac{4r}{r+1} - r \frac{F_\pi^2 M_\pi^2}{F_K^2 M_K^2}\\
&&
     -\frac{4r}{r+1} d_{K n} + r \frac{F_\pi^2 M_\pi^2}{F_K^2 M_K^2} d_{\pi n}
     +\frac{F_X^2 M_X^2}{F_K^2 M_K^2} d_{Xn}\,,\nonumber\\
\frac{F_X^2}{F_K^2} \label{fx}
  &=&2-\frac{F_\pi^2}{F_K^2}
      +\frac{1}{16\pi^2} \frac{M_\pi^2}{F_K^2} Y(n+2) \nonumber \\
      && \times
      \left[\frac{1}{2}\log\frac{M_\eta^2}{M_\pi^2} 
          + \frac{r}{n}\log\frac{M_\eta^2}{M_X^2} \right]
+\frac{F_X^2}{F_K^2} e_{Xn} \nonumber \\
&& + \frac{F_\pi^2}{F_K^2} e_{\pi n}
      -2 e_{K n} \,.
\end{eqnarray}
We have not discussed yet the identities for the $\eta$, which can 
be recasted in a form reminiscent of the Gell-Mann--Okubo formula:
\begin{eqnarray} \label{etanmass}
&&(n+2) F_\eta^2M_\eta^2 - 4 F_K^2M_K^2 - (n-2) F_\pi^2M_\pi^2 \nonumber \\
&&  =\quad 4(r-1)m^2 \left\{(r-1)(2n Z_n^p + A_n) \right.\nonumber \\
&& \left.\quad  + B^2_{0n}[(r-1)L'_n - L]\right\} +(n+2) F_\eta^2M_\eta^2d_\eta \nonumber \\
&& \quad- 4 F_K^2M_K^2 d_K - (n-2)
F_\pi^2M_\pi^2 d_\pi\,,\\
&&(n+2) F_\eta^2 - 4 F_K^2 - (n-2) F_\pi^2 \label{etandecay} \nonumber \\
&& =\quad \frac{2mB_{0n}}{16 \pi^2} 
  \left[\left(1+\frac{2r}{n}\right)
           \log\frac{M_\eta^2}{M_K^2} 
       - (2n-1) \log\frac{M_K^2}{M_\pi^2} \right. \nonumber \\
&& \left. \quad       
       + 2(n^2-1) \log\frac{M_X^2}{M_K^2}
  \right] + (n+2) F_\eta^2e_\eta - 4 F_K^2 e_K \nonumber \\
&& \quad  - (n-2) F_\pi^2 e_\pi\,.
\end{eqnarray}
The $\eta$-mass identity 
involves the new LEC $Z^p_n=16B_{0n}^2 L_7^n$. We have also introduced
the NNLO remainders $d_\eta$ and $e_\eta$.

The Ward identities are therefore expressions of $F_P^2$ and $F_P^2 M_P^2$ ($P=\pi,K,\eta,X$)
in terms of the fundamental parameters $r=m_s/m,\Sigma(n+2), F^2(n+2)$, the LEC's $L_{4 \ldots 8}$,
chiral logarithms of pseudoscalar masses and NNLO remainders.
The pion and kaon identities yield then $L_{4,5,6,8}$ in terms of $r,B_{0n},F^2(n+2)$,
chiral logarithms and NNLO remainders, see Eqs.~(\ref{L6}), (\ref{L8}), (\ref{L5}) and (\ref{L4}).  
We can obtain a similar expression for $L_7$ from the identity for $F_\eta^2M_\eta^2$. 
From Sec.~\ref{sec:connect}, pion and kaon Ward identities lead to the expression of
the parameters $B_{0n}$ and $F^2(n+2)$ as functions
of the two fluctuations parameters $u,v$, the ratio of quark masses $r$ and NNLO remainders 
[see Eqs.~(\ref{u})-(\ref{v})].

\begin{table}
\caption{Value of $M_X/M_K$ as a function of $Y(4)$ and $r$ for
$n=2$.\label{tab:mx}} 
\begin{center}
\begin{tabular}{|c|c|c|c|}
\hline
$Y(4)$ & $r=20$ & $r=25$ & $r=30$\\
\hline
0   & 1.43 & 1.37 & 1.30\\
0.2 & 1.44 & 1.38 & 1.31\\
0.5 & 1.44 & 1.38 & 1.31\\
1   & 1.46 & 1.40 & 1.32\\
2   & 1.48 & 1.42 & 1.35\\
4   & 1.57 & 1.53 & 1.46\\
\hline
\end{tabular}
\end{center}
\end{table}

We can now exploit all the remaining Ward identities 
to write $F_X^2$, $M_X^2$ and $F_\eta^2$ as functions
of $r,u,v$, chiral logarithms of $M_P$ and NNLO remainders. 
We need thus to know the masses $M_{\pi,K,\eta}$, the decay
constants $F_{\pi,K}$ and the NNLO remainders. As an illustration of
the method, we take the physical values of the masses and decay constants and 
set the remainders to zero, in order to obtain $M_X^2$
through an iterative procedure.
We start with the approximate value:
\begin{equation} \label{mxapprox}
\left.\frac{M_X^2}{M_K^2}\right|_{\rm start}=
  \frac{rF_K^2}{2F_K^2-F_\pi^2}
    \left[\frac{4}{r+1}-\frac{F_\pi^2 M_\pi^2}{F_K^2 M_K^2}\right]\,.
\end{equation}
and iterate Eqs.~(\ref{fxmx})-(\ref{fx}) until they converge to $M_X$. For instance,
the values for $M_X/M_K$ at $n=2$ are collected in Table~\ref{tab:mx}.

\section{Masses in the $SU(2)\times SU(2)$ chiral limit}
\label{app:fg}

To know the masses of the Goldstone bosons in the $SU(2)\times SU(2)$ chiral
limit, we take the limit $m\to 0$ in all the mass and decay constant
identities, and reexpress each LEC in this limit in terms of the same LEC with $m\neq 0$ and
chiral logarithms of $\bar{M}_P^2/M_P^2$, see Eqs.~(\ref{deltazs}) and (\ref{deltaxit}). Using
the previous relations, $\bar{F}_P^2$ and $\bar{M}_P^2$ are functions of $(r,u,v)$, 
pseudoscalar masses $M_Q^2$ and $\bar{M}_Q^2$, and remainders. If we keep on setting
$M_{\pi,K,\eta}$ and $F_{\pi,K}$ to their physical values, and neglecting NNLO remainders,
we can compute all the masses in the $SU(2)\times SU(2)$ limit in terms of the ratio of quark 
masses $r$ and the fluctuation parameters $u$ and $v$.

We need the pseudoscalar masses  $\bar{M}_P^2$
in particular to compute the factors $f_n$ and $g_n$
arising in the expression of $SU(2)\times SU(2)$ order parameters
Eqs.~(\ref{xtwo}) and (\ref{ztwo}). These factors can be written as:
\begin{equation}
\begin{array}{lcl}
\displaystyle
h_n = \frac{M_\pi^2}{F_\pi^2} \frac{r}{r-1} [L-L'_n]\,, &\quad &
\displaystyle
j = \frac{M_\pi^2}{F_\pi^2} 
  \frac{r+1}{r-1} \frac{1}{32\pi^2} \log\frac{M_\eta^2}{M_\pi^2}\,,\\
\displaystyle
b_n = -\frac{M_\pi^2}{F_\pi^2} \frac{nr+2}{2} \Delta Z^s_n\,, &\quad &
\displaystyle
c_n = -\frac{M_\pi^2}{F_\pi^2} (nr+2) \Delta \tilde\xi_n \,,\\
\displaystyle
f_n = h_n+b_n\,, &\quad &
\displaystyle
g_n =h_n+j+c_n\,.
\end{array}
\label{defbcfg}
\end{equation}
\begin{table}
\caption{Mass logarithms defined in Eq.~(\ref{defbcfg}) 
and involved in $Z(2)$ and $X(2)$, for
$r=25$, $n=1$ and $Z(3)=1$ ($j = 0.021$ and $h_1=0.059$). 
$M_{\pi,K,\eta}$ and $F_{\pi,K}$ are 
set to their physical values, and all 
NNLO remainders are neglected.\label{tab:chlogs}}
\begin{center}
\begin{tabular}{|c|cccc|}
\hline
$Y$ & $b$ & $c$ & $f$ & $g$\\
\hline
0 & -0.018 & -0.021 & 0.037 & 0.062\\
0.5 & -0.013 & -0.015 & 0.043 & 0.067\\
1 & -0.008 & -0.009 & 0.050 & 0.072\\
2 & -0.003 & -0.002 & 0.056 & 0.077\\
4 & -0.009 & -0.005 & 0.053 & 0.071\\
\hline
\end{tabular}
\end{center}
\end{table}
Table~\ref{tab:chlogs} shows an illustration of the physical case
$n=1$. The quark mass ratio is set to $r=25$.
$Z(3)$ and $Y(3)$ were chosen as parameters, rather than $u_1$ and $v_1$ 
[the two sets
are related through Eqs.~(\ref{u}) and (\ref{v})], and we set $Z(3)$ to 1.
$j$ is only a function of $r$. $h_n$ depends on $Y(n+2)$ only through $L'_n$, which vanishes
for $n=1$. For $r=25$, we have $j=0.021$ and $h_1=0.059$.
Furher study shows that $f_1$ and $g_1$ are only weakly dependent on
$Z$ and $r$. 
If we set the masses and decay constants to their physical values and
vary $n$, we observe that $f_n$ and $g_n$ are increasing functions of $n$.

\section{Large-$n$ limit} \label{app:ninf}

We take now the large-$n$ limit as described in Sec.~\ref{sec:8.1}. We stress
     that this limit is considered only at the level of the effective
     theory, with the additional assumption that the masses and decay
     constants tend to some particular asymptotic values. For
     numerical purposes, we will furthermore take these asymptotic
     values as the physical ones, and set the NNLO remainders to 0.
We recall that we have introduced in this limit two parameters $z_\infty$
and $a$ describing the behaviour of $SU(n+2)\times SU(n+2)$ chiral order parameters:
\begin{equation}
\begin{array}{l}
\displaystyle
Z(n+2) \to (1-\eta(r)-e_n) z_\infty\,,\\
\displaystyle
Y(n+2) \sim \frac{a}{n} \frac{1-\epsilon(r)-d_n}{1-\eta(r)-e_n}\,.
\end{array}
\end{equation}
The chiral logarithms disappear then from the mass identities 
Eqs. (\ref{pinmass})-(\ref{xnmass}). In the decay constant identities 
Eqs. (\ref{pindecay})-(\ref{xndecay}), only the logarithms of
$M_K^2/M_X^2$ survive -- $M_\eta$ disappears from the right hand side
of all mass and decay constant identities.
Due to this simplification, Eq.~(\ref{mxapprox}) becomes an exact formula for $M_X$ (up to NNLO
remainders). For $r=20,25,30$, $M_X/M_K$ is respectively 1.43, 1.37 and 1.30.

\begin{table}
\caption{Combination of chiral logarithms $l_\infty$ defined in Eq.~(\ref{linf}) and
involved in $Z(2)$ and $X(2)$ when $n\to\infty$.
$M_{\pi,K}$ and $F_{\pi,K}$ are set to their physical values, and all NNLO remainders are neglected.
\label{tab:chloginf}}
\begin{center}
\begin{tabular}{|c|ccc|}
\hline
$a$ & $r=20$ & $r=25$ & $r=30$\\
\hline
0 & 0.003 & 0.003 & 0.002\\
1 & 0.003 & 0.003 & 0.002\\
2 & 0.003 & 0.003 & 0.002\\
4 & 0.004 & 0.004 & 0.003\\
6 & 0.004 & 0.004 & 0.004\\
10 & 0.006 & 0.007 & 0.007\\
\hline
\end{tabular}
\end{center}
\end{table}

As far as the $\eta$-meson is concerned, the mass identity Eq.~(\ref{etanmass}) 
and the finiteness of all masses and decay constants at large $n$ result only in a constraint 
on the large-$n$ behaviour of $Z^p_n$ (i.e. $L_7^n$). On the other hand, the (LEC-free) 
relation for the decay constants Eq.~(\ref{etandecay}) imposes that $F_\eta=F_\pi$ up to NNLO 
remainders when $n\to\infty$. This can be related to the structure of the $\eta$ meson: 
$\lambda_\eta=\sqrt{1+2/n}\cdot {\rm diag}[1,1,-2/n \ldots -2/n]$, 
so that a similar equality may apply to $\pi$ and $\eta$ masses in this limit. 
Such an assumption is however not necessary for our purposes:
as we noticed earlier, the
large-$n$ behaviour of $M_\eta$ does not affect the $K$ and $X$-meson spectrum
since $\log M_\eta^2$ disappears from the Ward identities for the unmixed states.

Following the same lines as in the previous section, we can
exploit the Ward identities in the $SU(2)\times SU(2)$ chiral limit to determine
the pseudoscalar masses for $m\to 0$. Let us in particular notice the identities for
the kaons:
\begin{eqnarray} \label{barkdecay}
\frac{\bar{F}_K^2}{F_\pi^2}
 &\sim & 1+\left(\frac{r}{2}-1\right)\eta+ \frac{r}{2} e'_n -e_n +
\frac{\bar{F}_K^2}{F_\pi^2}\bar{e}_{Kn} \nonumber \\  
&&  +\frac{ra}{64\pi^2}\frac{1-\epsilon(r)-d_n}{1-\eta(r)-e_n}\frac{M_\pi^2}{F_\pi^2}
  \left[2\log\frac{M_X^2}{\bar{M}_X^2} \right. \nonumber \\
&& \left.-\frac{2}{r-1}\log\frac{M_X^2}{M_K^2}
     -\log\frac{\bar{M}_K^2}{M_K^2} \right] \,,\\
\frac{\bar{F}_K^2 \bar{M}_K^2}{F_\pi^2 M_\pi^2} &=&
  \frac{r}{2}\left[1+\left(\frac{r}{2}-1\right)\epsilon +\frac{r}{2}d'_n - d_n\right]
  +\frac{\bar{F}_K^2 \bar{M}_K^2}{F_\pi^2 M_\pi^2}\bar{d}_{Kn}\,,\nonumber \\
  &&
\end{eqnarray}
and the extra-$X$ states:
\begin{eqnarray} \label{barxdecay}
\frac{\bar{F}_X^2}{F_\pi^2}
 &\sim & \frac{F_K^2}{F_\pi^2}\left[2-\frac{F_\pi^2}{F_K^2}\right]
   - \frac{ra}{16\pi^2} 
   \frac{1-\epsilon(r)-d_n}{1-\eta(r)-e_n}\frac{M_\pi^2}{F_\pi^2} 
    \log\frac{\bar{M}_X^2}{M_X^2} \nonumber \\
 && + re'_n - e_n +  \frac{\bar{F}_X^2}{F_\pi^2} \bar{e}_{Xn}\,,
 \end{eqnarray}
 \begin{eqnarray}
\frac{\bar{F}_X^2 \bar{M}_X^2}{F_\pi^2 M_\pi^2} &=&
  \frac{F_K^2M_K^2}{F_\pi^2 M_\pi^2}
  \left[\frac{4r}{r+1}
    - r \frac{F_\pi^2 M_\pi^2}{F_K^2 M_K^2}\right]+r(rd'_n - d_n) \nonumber \\
 && + 
 \frac{\bar{F}_X^2 \bar{M}_X^2}{F_\pi^2 M_\pi^2}\bar{d}_{Xn}\,.\label{barxmass}
\end{eqnarray}
Comparing Eqs.~(\ref{barxdecay})-(\ref{barxmass}) with Eqs.~(\ref{fxmx})-(\ref{fx})
shows that $\bar{F}_X^2=F_X^2$ and $\bar{M}_X^2=M_X^2$ in the 
large-$n$ limit (up to NNLO remainders).

The chiral logarithms involved in the discussion of $X(2)$ and $Z(2)$ reduce
now to a single combination of logarithms, independent of $\eta$ observables:
\begin{eqnarray} \label{linf}
l_\infty &=& \lim_{n\to\infty} \frac{f_n}{n} 
  = \lim_{n\to\infty} \frac{g_n}{n} \\
  &=& \lim_{n\to\infty} \frac{M_\pi^2}{F_\pi^2} \frac{r}{32 \pi^2 }
\left[\frac{2}{r-1}\log\frac{M_X^2}{M_K^2}+\log\frac{\bar{M}_K^2}{M_K^2}
\right] \,. \nonumber \\
&&
\end{eqnarray}

If we neglect (as a first approximation) the chiral logarithms in Eq.~(\ref{barkdecay}) -- 
setting $a=0$ -- we obtain:
\begin{eqnarray} \label{linfa0}
\left. l_\infty \right|_{a=0}
&=&\frac{M_\pi^2}{F_\pi^2} \frac{r}{32\pi^2 }\Bigg\{
 \log \frac{r}{2}\frac{F_\pi^2M_\pi^2}{F_K^2 M_K^2}
 +\log \frac{1+(r/2-1)\epsilon(r)}{1+(r/2-1)\eta(r)} \nonumber \\
&&
  +\frac{2}{r-1}
\left[\log\left(4-\frac{4}{r+1}-r\frac{F_\pi^2M_\pi^2}{F_K^2 M_K^2}\right)
\right. \nonumber \\
&& \left.
  -\log\left(2\frac{F_\pi^2}{F_K^2}-1\right)\right]\Bigg\}\,. 
\end{eqnarray}

In Eq.~(\ref{linfa0}),
we would na\"{\i}vely expect the first line 
to be dominant at large $r$. This is not actually the case. We have:
\begin{equation}
\log \frac{r}{2}\frac{F_\pi^2M_\pi^2}{F_K^2 M_K^2}
=\log \frac{r}{r^*_2+1}\,,\qquad
r^*_2 = 2\frac{M_K^2}{M_\pi^2}-1=24\,,
\end{equation}
and $(\epsilon-\eta)(r)$ is precisely changing sign at $r^*_2$, being positive for $r<r^*_2$
and negative $r>r^*_2$. We see therefore that the two logarithms in the first line
of Eq.~(\ref{linfa0}) are of opposite signs, exchanging them for $r\sim 24$. 

The actual sign of $l_\infty$ is therefore a question of subtle compensation between all the logarithms
involved in its expression. Following the procedure outlined above,
we have computed the values of $l_\infty$ for various values of $r$ and $a$,
collected in Table~\ref{tab:chloginf}, with
$M_{\pi,K}$ and $F_{\pi,K}$ set to their physical values,
and all NNLO remainders neglected. We see that $l_\infty$ remains
small and positive in any case.

\end{document}